\documentclass[pra,aps,showpacs,nofootinbib,floatfix,11pt]{revtex4-2}
\usepackage{amsfonts}
\usepackage{amssymb}
\usepackage{mathrsfs}
\usepackage{amsthm}
\usepackage{graphicx}
\usepackage{amsmath}
\usepackage[colorlinks=true,citecolor=blue,urlcolor=blue]{hyperref}
\usepackage{color}
\begin{document}
%
%\title{When does non-Gaussianity help? Characterization of nonlinear media using coherent and squeezed probes with photon addition and subtraction}

\title{Squeezing as a catalyst for non-Gaussian advantage in characterization of nonlinear media}

\author{Manju$^{1,2,5}$, Samaneh Hesabi$^{3}$, Siting Tang$^{4,5}$, Peter van Loock$^{3}$, Matteo G. A. Paris$^{5}$}
\affiliation{\(^1\)Harish-Chandra Research Institute, Chhatnag Road, Jhunsi, Allahabad - 211019, India\\
\(^2\) Homi Bhabha National Institute, Training School Complex, Anushakti Nagar, Mumbai 400 094, India}
\affiliation{$^3$Institute of Physics, Johannes Gutenberg-Universität Mainz, Staudingerweg 7, 55128 Mainz, Germany}
\affiliation{$^4$ State Key Laboratory of Mathematical Sciences, Academy of Mathematics and Systems Science, Chinese Academy of Sciences, Beijing 100190, China}
%\affiliation{$^4$ School of Mathematical Sciences, University of Chinese Academy of Sciences, Beijing 100049, China}
\affiliation{$^5$ Dipartimento di Fisica dell'Università di Milano, I-20133 Milan, Italy}
%\affiliation{$^6$ INFN, Sezione di Milano, I-20133 Milan, Italy}
\date{\today}

\begin{abstract}
We address the precise characterization of coupling strength of nonlinear media in continuous-variable (CV) quantum systems using coherent and squeezed vacuum states as Gaussian probes, together with their photon-added and photon-subtracted counterparts as non-Gaussian probes. We consider three main classes of nonlinear Hamiltonians, namely quadrature nonlinearities, generalized squeezing and Kerr-type interactions. By analytically evaluating the quantum Fisher information (QFI), we compare the performance of Gaussian and non-Gaussian probes and assess the optimal probe based on the probe parameters, energy resource and non-Gaussianity. Our results are twofold as follows: first, for coherent-state family, the improvement provided by photon addition at fixed coherent amplitude originates mainly from the extra energy carried by the probe and does not provide a genuine metrological resource, since the same precision can be achieved by a Gaussian coherent-state signal of a larger energy, which can be more easily produced. Second, in contrast, photon addition and subtraction become effective resources when applied to already nonclassical states such as squeezed vacuum states. In this case, they lead to a significant enhancement of the QFI, particularly for higher-order interactions. Although Gaussian squeezed states remain optimal at equal energy constraint, photon-added and photon-subtracted squeezed states achieve comparable sensitives with significantly lower squeezing requirements. Since large squeezing level remains experimentally challenging, these non-Gaussian probes offer a practical route towards enhanced estimation of the nonlinear coupling strength within currently accessible squeezing regimes. 
%Our results highlights the beneficial interplay between non-Gaussian state engineering and nonlinearity parameter estimation, and identify de-Gaussified squeezed states as promising probes for the characterization of nonlinear quantum media.
\end{abstract}
\maketitle
\section{Introduction}
Quantum estimation theory provides the fundamental framework for determining unknown physical parameters with the highest precision allowed by quantum mechanics \cite{Helstrom1969,doi:10.1142/S0219749909004839}. It underpins a broad range of quantum technologies, including quantum sensing, metrology, communication, and computation, all of which rely on the accurate characterization and control of physical parameters \cite{doi:10.1126/science.1104149,PhysRevLett.96.010401,RevModPhys.89.035002}. The precision achievable in any estimation protocol is bounded by the quantum Cram{\'e}r-Rao inequality \cite{cramer1999mathematical}, where QFI quantifies the sensitivity of a probe state to variations of the parameter of interest. Consequently, identifying probe states that maximize the QFI is therefore a key objective in quantum metrology.

%In quantum information science, quantum estimation is crucial as it offers a systematic approach to accurately determining physical parameters using quantum systems \cite{matteo1}. In contrast to traditional estimation techniques, which are limited by the inherent constraints of the uncertainty principle \cite{cramer}, quantum estimation utilizes the principles of quantum mechanics to achieve unparalleled precision in measurements. Central to this framework is the concept of Quantum Fisher Information (QFI), which serves as a metric for quantifying the information content of a quantum state with respect to the parameter of interest \cite{hel}. Quantum estimation encompasses a diverse array of applications, including quantum metrology \cite{gio,esc}, quantum sensing \cite{deg}, and other quantum protocols \cite{yang,hu2}.
Among the various physical platforms employed for quantum-enhanced estimation, CV systems have emerged as a highly versatile and experimentally accessible platform for quantum technologies, supported by their mature optical implementations \cite{RevModPhys.77.513,doi:10.1142/p489}.  In these systems, Gaussian states such as coherent and squeezed vacuum states play a central role due to their efficient generation, manipulation, and detection \cite{ferraro2005gaussianstatescontinuousvariable,RevModPhys.84.621}. In particular, squeezed states enable sensitivities beyond classical limits and constitute key resources for quantum-enhanced metrology \cite{PhysRevD.23.1693,PhysRevLett.59.278,GAIBA2009934,Aasi2013,app151810179,PhysRevResearch.6.033292, https://doi.org/10.1002/qute.202501012}. However, Gaussian operations and states alone are insufficient for universal CV quantum computation. Universal CV quantum computation requires the inclusion of non-Gaussian operations generated by higher-order nonlinear interactions, such as Kerr interactions and cubic phase gates \cite{PhysRevLett.82.1784,PhysRevA.65.042304, PhysRevA.64.012310}. These nonlinear operations enable functionalities inaccessible within the Gaussian framework \cite{PhysRevA.104.032420,PhysRevA.102.012608,PhysRevResearch.6.023332}. Despite their importance, the practical implementation of such nonlinear processes remains extremely challenging because their performance is highly sensitive to the precise strength and control of the underlying nonlinear interactions \cite{PhysRevA.91.032321,PhysRevA.93.022301,PhysRevLett.125.160501,PhysRevApplied.15.024024,Sakaguchi2023}. In particular, even small uncertainties in nonlinear couplings can significantly affect the fidelity of Kerr evolutions, cubic phase operations, and other non-Gaussian transformations required for universal CV quantum computation \cite{PhysRevA.84.053802,2q95-sfjs}. In this context, the precise estimation of coupling strengths of nonlinear media becomes fundamentally important. Developing optimal strategies for estimating these coupling strengths is therefore a crucial step toward scalable and controllable CV quantum technologies. However, unlike linear interaction, nonlinear quantum evolutions encode information through higher-order moments, intricate phase space structures, and strongly nonclassical correlations \cite{Agarwal_2005,PhysRevA.76.042327,PhysRevLett.124.063605,verma2026nongaussianentanglementrevealedhigherorder} that cannot be fully captured within the Gaussian framework alone. As a result, non-Gaussian states are expected to provide enhanced sensitivity and improved estimation of coupling strength of nonlinear quantum media. Among the various non-Gaussian resources, photon-added and photon-subtracted states are particularly attractive because they can be experimentally generated through de-Gaussification protocols applied to Gaussian states \cite{PhysRevA.43.492,PhysRevLett.92.153601,Kim_2008,PhysRevLett.101.233605,PhysRevA.82.031802,PhysRevA.82.063833,doi:10.1126/science.1146204,Pasharavesh_2024}.
%These operations significantly modify the photon number distribution and phase space structure of the original states, leading to enhanced nonclassicality and richer quantum correlations.
Such states have already demonstrated advantages in several quantum information and quantum metrology tasks \cite{PhysRevA.85.013839,PhysRevA.87.052321,PhysRevA.87.022313,PhysRevLett.93.130409,PhysRevA.90.013821,PhysRevA.86.012328,e23101353,PhysRevA.101.063810,PhysRevA.111.022402,PhysRevA.111.043719,doi:10.1142/S0219477525400334}.

Motivated by these facts, in this work, we investigate the precise estimation of coupling strength of nonlinear media. Specifically, we consider the estimation of the parameter $\lambda$ encoded through the unitary evolution $U_{\lambda} = \exp(-i\lambda \hat{G})$, where $\hat{G}$ represents different classes of linear and nonlinear generators. We focus on three relevant families of generators given by $(\hat{a}+\hat{a}^{\dagger})^n$, $\hat{a}^n+\hat{a}^{\dagger\,n}$, and $\hat{a}^{\dagger\,n}\,\hat{a}^n$, considering nonlinear orders $n=1,2,3,4$. These generators describe quadrature nonlinearities, generalized squeezing interactions, and Kerr-type nonlinear processes, respectively. As probe states, we employ coherent and squeezed vacuum states as Gaussian resources together with their non-Gaussian counterparts generated via photon addition and subtraction. By evaluating and comparing the QFI associated with these probes, we analyze the role of probe parameters, average energy, and degree of non-Gaussianity (NG) in the estimation process. Our goal is to identify the optimal probe for nonlinear estimation. The results reveal a clear distinction between the metrological roles of de-Gaussification in coherent and squeezed state families. For coherent probes, the enhancement achieved through photon addition can largely be reproduced by increasing the energy of coherent state, indicating that NG does not provide a genuine metrological advantage in this case. In contrast, photon-added and photon-subtracted squeezed states exhibit substantially enhanced sensitivities that cannot be efficiently reproduced by squeezed states without requiring unrealistically large squeezing strengths.\\ 
The remainder of the paper is organized as follows. In Sec. \ref{sec:II}, we briefly review the essentials of local quantum estimation theory and the definition of the QFI. Section \ref{sec:III} introduces the Gaussian and non-Gaussian probe states employed throughout this work. In Sec. \ref{sec:IV}, we derive the general expressions for the QFI corresponding to the nonlinear Hamiltonians under consideration. Sections \ref{sec:V} and \ref{sec:VI} present a detailed analysis of the estimation performance and identification the optimal probes under different criteria. Finally, Sec. \ref{sec:VII} summarizes our conclusions and discusses the implications of our results for nonlinear quantum metrology in CV systems.

\section{Local Quantum Estimation Theory}
\label{sec:II}
The primary objective of the estimation theory is to determine the value of an unknown parameter with the highest possible precision from a set of experimental observations. In the context of classical estimation theory, to estimate the value of a parameter $\lambda$, we aim to find an estimator, denoted as $\hat{\lambda
 }:\Gamma \to \mathcal{M}_{\lambda}$, where $\Gamma=(\gamma_{1},\gamma_{2},...)$ represents the experimental data and $\mathcal{M}_{\lambda}$ is the set of possible values of parameter $\lambda$. The Cram{\'e}r-Rao bound in classical estimation theory sets the fundamental limit on the precision attainable in parameter estimation, and the most effective estimators are those that meet this bound \cite{HELSTROM1967101,Helstrom1969,Rao1992,doi:10.1142/S0219749909004839}:
\begin{equation}\label{crR}
V(\hat{\lambda})\geq \frac{1}{M F(\lambda)}
\end{equation}

The expression $V(\hat{\lambda})=E_{\lambda}[(\hat{\lambda}({\gamma})-\lambda)^2]$ represents the mean square error of an estimator for the parameter $\lambda$,  where $M$ denotes the number of measurements and $F(\lambda)$ is the classical Fisher information (CFI), defined as,
\begin{equation}\label{fish}
F(\lambda)=\int d\gamma \, p(\gamma| \lambda) \bigg(\frac{\partial \,ln\, p(\gamma|\lambda)}{\partial \lambda}\bigg)^{2}=\int d\gamma 
\frac{1}{ p(\gamma| \lambda)}\bigg(\frac{\partial \, p(\gamma| \lambda)}{\partial \lambda}\bigg)^{2}
\end{equation}
where $p(\gamma| \lambda)$ denotes the probability distribution of obtaining the measurement outcome $\gamma$ when the parameter has the value $\lambda$ \cite{Helstrom1969,Rao1992}.

The quest for precision in quantum mechanics goes beyond the constraints of classical limitations, and the quantum Cram{\'e}r-Rao bound offers a theoretical framework for comprehending the ultimate boundaries of this pursuit. Within the realm of quantum mechanics, conventional CFI is replaced by the concept of QFI, which serves as a critical metric for revealing the sensitivity of a quantum state to parameter variations.

In a situation with only one parameter, optimization can be improved by taking into account all potential measurements, resulting in the development of a quantum Cram{\'e}-Rao inequality for a single parameter. The QFI is defined as \cite{doi:10.1142/S0219749909004839,Liu_2016}

\begin{equation}\label{QFI}
\mathcal{F}(\lambda) = \text{Tr}\left[\rho_{\lambda}L_{\lambda}^{2}\right],
\end{equation}

where $\rho_{\lambda}$ is the family of quantum states that depend on $\lambda$, and $L_\lambda$ is the symmetric logarithmic derivative (SLD) which is a Hermitian operator that quantifies how changes in a quantum state affect a parameter of interest. It is defined by the Lyapunov equation as 
\begin{equation}\label{SLD} 
\frac{\partial \rho_{\lambda}}{\partial \lambda}=\frac{1}{2}\{L_{\lambda},\rho_{\lambda}\}
\end{equation}  
In the case of pure states, $\rho_{\lambda}=\vert\psi_{\lambda}\rangle\langle\psi_{\lambda}\vert$, since $\rho_{\lambda}^2=\rho_{\lambda}$, we have $\partial_{\lambda} \rho_{\lambda}=\partial_{\lambda} \rho_{\lambda}\rho_{\lambda}+\rho_{\lambda}\partial_{\lambda}  \rho_{\lambda}$ and therefore the SLD operator is abbreviated as $L_{\lambda}=2\partial_{\lambda} \rho_{\lambda}$. Ultimately, the QFI is simplified as \cite{PhysRevLett.72.3439}
\begin{equation}\label{purest}
\mathcal{F}(\lambda) = 4\big[\langle\partial_{\lambda}\psi_{\lambda}\vert\partial_{\lambda}\psi_{\lambda}\rangle- |\langle \partial_{\lambda}\psi_{\lambda}\vert \psi_{\lambda}\rangle|^2\big]
\end{equation}
In this study, our focus will be on systems in which the connection between the parameter $\lambda$ and the density operator $\rho_{\lambda}$ is generated by a family of unitary transformations, denoted as $\rho_{\lambda}=U_{\lambda}\rho_{0}U^{\dagger}_{\lambda}$, where $U_{\lambda}=\exp(-i\lambda \hat{G})$, with $\hat{G}$ representing the generator responsible for the transformation, and $\rho_{0}$ is a specific quantum state employed to probe the Hamiltonian process.
In this case,  it is possible to derive a specific equation for QFI.
If the initial probe is assumed to be in a pure state $\vert\psi_0\rangle$, then the resulting state after evolution, $\vert\psi_\lambda\rangle$, will also be pure. Hence, the QFI derived from Eq. (\ref{purest}), can be expressed in terms of the generator $\hat{G}$ and the initial probe as \cite{doi:10.1142/S0219749909004839} as
\begin{equation}\label{purecase}
\mathcal{F}(\lambda) =  4\left[\langle\psi_0|\hat{G}^{2}|\psi_0\rangle -\langle\psi_0|\hat{G}|\psi_0\rangle^2\right]\
\end{equation}

\section{Classes of Gaussian and non-Gaussian Probes}
\label{sec:III}
%Continuous variable quantum system provide an important platform for quantum estimation and quantum metrology. In this framework, gaussian states such as coherent and squeezed states are widely employed because of their experimental accessibility and simple mathematical structure. However, non-Gaussian states possess enhanced nonclassical properties that may improve the estimation precision of considered system. In particular, photon addition and photon subtraction operations are standard de-Gaussification protocols capable of generating non-Gaussian resource from Gaussian seed states. In this work, we consider coherent states and squeezed vacuum states as Gaussian probes, while their photon-added and photon-subtracted counterparts are employed as non-Gaussian probes. 

In this section, we introduce the probe states employed for the estimation of $\lambda$. We consider both Gaussian and non-Gaussian CV states in order to analyze how de-Gaussification affects the achievable estimation precision. The Gaussian probes considered in this work are coherent states (CSs) and squeezed vacuum states (SSs). A CS is defined as
\begin{equation}\label{coh}
    |\alpha\rangle=\hat{D}(\alpha)\,|0\rangle,
\end{equation}
where $\hat{D}(\alpha)=e^{\alpha \hat{a}^{\dagger}-\alpha^*\hat{a}}$ is the single-mode displacement operator, and $\alpha=|\alpha| e^{i\phi}$ denotes the coherent amplitude. Similarly, the SS is given by 
\begin{equation}\label{squ}
    |\xi \rangle= \hat{S}(\xi)\,|0\rangle,
\end{equation}
where $\hat{S}(\xi)=\exp\left[\frac{1}{2}\xi\hat{a}^{\dagger\,2 }-\frac{1}{2}\xi^* \hat{a}^2\right]$ is the single-mode squeezing operator, with $\xi=r e^{i\theta}$, where $r$ is the squeezing strength and $\theta$ is the squeezing phase, respectively. For convince, we define $\nu=\sinh{r}e^{i\theta}$, such that $|\nu|^2=\sinh^{2}{r}$. The non-Gaussian probes are obtained by applying photon addition and subtraction operations to these Gaussian states. The normalized photon-added coherent state (PACS) \cite{PhysRevA.43.492} and photon-subtracted coherent state (PSCS) \cite{Zavatta_2008} are defined as
\begin{equation}\label{chadd}
  |\alpha \rangle_{add}=\frac{\hat{a}^{\dagger}|\alpha\rangle}{\sqrt{1+|\alpha|^2}},\,\,\,\,\,\,\,\,\,\,\,\,|\alpha  \rangle_{sub}=\frac{\hat{a}|\alpha  \rangle}{\sqrt{|\alpha|^2}}
\end{equation}
Likewise, the normalized photon-added squeezed state (PASS) \cite{ZHANG199214} and photon-subtracted squeezed state (PSSS) \cite{PhysRevLett.92.153601} are given by
\begin{equation}\label{sqadd}
  |\xi \rangle_{add}=\frac{\hat{a}^{\dagger}|\xi\rangle}{\sqrt{1+|\nu|^2}},\,\,\,\,\,\,\,\,\,\,\,\,|\xi  \rangle_{sub}=\frac{\hat{a}|\xi  \rangle}{\sqrt{|\nu|^2}}
\end{equation}
An important distinction appears between coherent and squeezed probes under photon subtraction for the $\lambda$ estimation. Since, CSs are eigenstates of annihilation operator, $\hat{a}|\alpha\rangle=\alpha|\alpha\rangle$, the PSCS reduces to the original CS \cite{Zavatta_2008,PhysRevA.98.013809}. Therefore, PSCSs do not represent genuinely new non-Gaussian resources and possess the same physical properties and the same QFI as CSs. However, the situation is fundamentally different for SSs. Unlike CSs, SSs contain only even photon number components in the Fock basis \cite{10.1093/acprof:oso/9780198563617.001.0001,10.1093/oso/9780198501770.001.0001}. Therefore, the action of either the creation operator $\hat{a}^\dagger$ or the annihilation operator $\hat{a}$ transforms the even parity SS into an odd parity state with closely related photon number statistics. Consequently, expectation values entering the QFI expressions become identical for PASSs and PSSSs. As a result, these two classes of probes yield the same QFI for all nonlinear generators considered in this work. This equivalence explains that photon addition and subtraction on SSs exhibit identical metrological performances throughout our analysis. Thus, in this work, we employ CSs, PACSs, SSs, and PASSs as probes. Having introduced the different probes considered in this work, we now investigate their performance in parameter estimation. In the following section, we evaluate the QFI associated with these probes for different Hamiltonians.

\section{QFI of different Nonlinear Generators}
\label{sec:IV}
In this Section, we evaluate the QFI associated with the estimation of coupling for different generators. More specifically, we provide the QFI of Hamiltonians,  $\hat{H}_{x}=\lambda(\hat{a}+\hat{a}^{\dagger})^n$, $\hat{H}_{s}=\lambda(\hat{a}^n+\hat{a}^{\dagger\,n})$, and $\hat{H}_{k}=\lambda(\hat{a}^{\dagger\,n}\,\hat{a}^n)$.
\subsection{Hamitonian $\hat{H}_x= \lambda\,\left(\hat{a}+\hat{a}^{\dagger}\right)^n$}
For this Hamiltonian, the QFI for each probe $|\psi_{0}\rangle$, is given by
\begin{equation}\label{purecase1}
\mathcal{F}(\lambda) =  4\left[\langle\psi_0|(\hat{a}+\hat{a}^{\dagger})^{2n}|\psi_0\rangle -\langle\psi_0|(\hat{a}+\hat{a}^{\dagger})^n|\psi_0\rangle^2\right]\
\end{equation}
To calculate the QFI, we need to obtain the expectation values of $(\hat{a}+\hat{a}^{\dagger})^{2n}$ and $(\hat{a}+\hat{a}^{\dagger})^{n}$ for different states. Considering that the creation and annihilation operators satisfy $[\hat{a},\hat{a}^\dagger]=\mathbb{I}$, the normal ordering 
of $(\hat{a}+\hat{a}^{\dagger})^n$ may be written as \cite{10.1063/1.1705306, Gerry01061993, osti_4208943}.
\begin{equation}\label{ex}
\begin{array}{cc}
(\hat{a}+\hat{a}^\dagger)^n = \sum\limits_{m=0}^{\lfloor n/2 \rfloor} \sum\limits_{s=0}^{n-2m} \frac{n!}{2^m s!m! (n-s-2m)!}\hat{a}^{\dagger\,s} \hat{a}^{n-s-2m},
\end{array}
\end{equation}
This equation is applicable for calculating expectation values for states that maintain the normal order of operators. Nevertheless, for photon added states, $|\psi_{1}\rangle=\hat{a}^\dagger|\psi_{0}\rangle$, the expectation value takes the following form:
\begin{equation}\label{ex1}
\begin{array}{cc}
\langle \psi_{1}| (\hat{a}+\hat{a}^\dagger)^n | \psi_{1}\rangle=
\sum\limits_{m=0}^{\lfloor n/2 \rfloor} \sum\limits_{s=0}^{n-2m} \frac{n!}{2^m s!m! (n-s-2m)!}\,
\langle \psi_{0}|\hat{a}\hat{a}^{\dagger\,s} \hat{a}^{n-s-2m}\hat{a}^\dagger | \psi_{0}\rangle\,,
\end{array}
\end{equation}
and the calculation of the expectation values is less straightforward. To simplify this expression we use  the normal order formula \cite{Gerry01061993, osti_4208943} 
\begin{equation}\label{ord}
\hat{a}^{k}\hat{a}^{\dagger\,l}=\sum_{m=0}^{\min[k,l]} m! \binom{k}{m} \binom{l}{m}\hat{a}^{\dagger\,l-m}\hat{a}^{k-m}\,,
\end{equation}
such that Eq. (\ref{ex1}) may be rewritten as 
\begin{align}\label{exx}
&\langle \psi_{1}| (\hat{a}+\hat{a}^\dagger)^n |\psi_{1}\rangle
=\sum\limits_{m=0}^{\lfloor n/2 \rfloor} \sum\limits_{s=0}^{n-2m}\sum\limits_{p=0}^{\min[1,s]}\sum\limits_{q=0}^{\min[1,n-s-2m]} \sum\limits_{j=0}^{\min[1-p,1-q]} \binom{1}{p}\binom{s}{p}
\binom{1}{q} \binom{n-s-2m}{q} \notag \\
&\times \binom{1-p}{j}\binom{1-q}{j}
\frac{n! p! q! j!}{2^m s!m! (n-s-2m)!}\,
\langle \psi_{0}|\hat{a}^{\dagger\,s-p-q-j+1} \hat{a}^{n-s-2m-p-q-j+1}|\psi_{0}\rangle
\end{align}
\subsection{Hamitonian $\hat{H_s}= \lambda(\hat{a}^n+\hat{a}^{\dagger\,n})$}
For this Hamiltonian, the QFI for each probe, $|\psi_{0}\rangle$, is given by
\begin{equation}\label{H2}
\mathcal{F}(\lambda) =  4\left[\langle\psi_0|(\hat{a}^n+\hat{a}^{\dagger\,n})^{2}|\psi_0\rangle -\langle\psi_0|(\hat{a}^n+\hat{a}^{\dagger\,n})|\psi_0\rangle^2\right]\
\end{equation}

After extending the first part and using Eq. (\ref{ord}), we have 
\begin{equation}\label{H22}
\mathcal{F}(\lambda) =  4\left[\langle\psi_0|\left(\hat{a}^{2n}+\hat{a}^{\dagger\,2n}+\hat{a}^{\dagger\,n}\hat{a}^{n}+\sum_{q=0}^n q!\binom{n}{q}\binom{n}{q}\hat{a}^{\dagger\,n-q} \hat{a}^{n-q}\right)|\psi_0\rangle -\langle\psi_0|(\hat{a}^n+\hat{a}^{\dagger\,n})|\psi_0\rangle^2\right]\
\end{equation}
For PAs, $|\psi_{1}\rangle=\hat{a}^{\dagger}| \psi_{0}\rangle$, the QFI is given by
\begin{align}\label{H222}
\mathcal{F}(\lambda) = & 4\Bigg[\langle\psi_0|\left(\hat{a}^{2n+1}\hat{a}^{\dagger}+\hat{a}\hat{a}^{\dagger\,2n+1}+\hat{a}\hat{a}^{\dagger\,n}\hat{a}^{n}\hat{a}^{\dagger}+\sum_{q=0}^n q!\binom{n}{q}\binom{n}{q}\hat{a}\hat{a}^{\dagger\,n-q} \hat{a}^{n-q}\hat{a}^{\dagger}\right)|\psi_0\rangle \nonumber \\
& \quad -\langle\psi_0|(\hat{a}^{n+1}\hat{a}^{\dagger}+\hat{a}\hat{a}^{\dagger\,n+1})|\psi_0\rangle^2\Bigg]\
\end{align}
Now, we apply Eq. (\ref{ord}) to rewrite all parts of Eq. (\ref{H222}) in the normal order and we have 
\begin{align}
\mathcal{F}(\lambda) =& 4\Bigg[ \langle\psi_0|\Bigg(\sum_{f=0}^{1} f!\binom{2n+1}{f}\binom{1}{f}\hat a^{\dagger(1-f)} \hat a^{2n+1-f} +\sum_{g=0}^{1}g!\binom{1}{g}\binom{2n+1}{g} \hat a^{\dagger(2n+1-g)}\hat a^{1-g} \nonumber\\
& +\sum_{h=0}^{1}\sum_{i=0}^{1}\sum_{j=0}^{\min(1-h,1-i)}h!i!j!\binom{1}{h}\binom{n}{h}\binom{n}{i}\binom{1}{i} \binom{1-h}{j}\binom{1-i}{j} \hat a^{\dagger(n+1-h-i-j)} \hat a^{\,n+1-h-i-j}\nonumber\\
& +\sum_{q=0}^{n}p!q!r!s!\binom{n}{q}^{2}\sum_{p=0}^{1}\sum_{r=0}^{1}\sum_{s=0}^{\min(1-p,1-r)} \binom{1}{p}\binom{n-q}{p} \binom{n-q}{r}\binom{1}{r}\binom{1-p}{s}\binom{1-r}{s}\nonumber\\
&\qquad \hat a^{\dagger(n-q+1-p-r-s)}\hat a^{\,n-q+1-p-r-s} \Bigg)|\psi_0\rangle - \Bigg(\langle\psi_0|\sum_{k=0}^{1}k!\binom{n+1}{k}\binom{1}{k}\hat a^{\dagger(1-k)}\hat a^{\,n+1-k}\nonumber\\
& \qquad + \sum_{l=0}^{1}l!\binom{1}{l}\binom{n+1}{l}\hat a^{\dagger(n+1-l)}\hat a^{\,1-l}|\psi_0\rangle\Bigg)^2\Bigg].
\end{align}

%For example, in the case of PSs, $|\psi_{2}\rangle=\hat{a}| \psi_{0}\rangle$, the QFI is given by
%\begin{align}
%\label{H223}
% \mathcal{F}(\lambda) =&  4\Bigg\{\langle \psi_{0}|\hat{a}^{\dagger}\hat{a}^{2n+1}+\hat{a}^{\dagger\,2n+1} \hat{a}+\hat{a}^{\dagger\,n+1}\hat{a}^{n+1}+\sum_{q=0}^{n} q!\binom{n}{q}\binom{n}{q}\hat{a}^{\dagger\,n-q+1} \hat{a}^{n-q+1}| \psi_{0}\rangle \nonumber \\
%& \quad -\langle \psi_{0}|\hat{a}^{\dagger}\hat{a}^{n+1}+\hat{a}^{\dagger\,n+1}\hat{a}| \psi_{0}\rangle^2\Bigg\}   
%\end{align}

\subsection{Hamitonian $\hat{H_k}=\lambda(\hat{a}^{\dagger\,n} \hat{a}^n)$}
For this Hamiltonian, the QFI has this form: 
\begin{equation}\label{H3}
\mathcal{F}(\lambda) =  4\left[\langle\psi_0|(\hat{a}^{\dagger\,n}\,\hat{a}^n)^{2}|\psi_0\rangle -\langle\psi_0|(\hat{a}^{\dagger\,n}\,\hat{a}^n)|\psi_0\rangle^2\right]
\end{equation}

In normal order, this can be written as 
\begin{equation}\label{H31}
\mathcal{F}(\lambda) =4\left[\langle\psi_0|\sum_{p=0}^{n}p!\binom{n}{p}\binom{n}{p}\hat{a}^{\dagger\,2n-p} \hat{a}^{2n-p}|\psi_0\rangle -\langle\psi_0|\hat{a}^{\dagger\,n} \hat{a}^{n}|\psi_0\rangle^2\right]\
\end{equation}

For PAs, $|\psi_{1}\rangle=\hat{a}^\dagger|\psi_{0}\rangle$, the QFI has the following form:
\begin{equation}\label{H32}
\mathcal{F}(\lambda) =4\left[\langle\psi_0|\sum_{p=0}^{n}p!\binom{n}{p}\binom{n}{p}\hat{a}\hat{a}^{\dagger\,2n-p} \hat{a}^{2n-p}\hat{a}^{\dagger}|\psi_0\rangle -\langle\psi_0|\hat{a}\hat{a}^{\dagger\,n} \hat{a}^{n}\hat{a}^{\dagger}|\psi_0\rangle^2\right]\
\end{equation}

Using Eq. (\ref{ord}), the final form of QFI is given by 
%\begin{equation}\label{H32}
%\begin{aligned}
    %\mathcal{F}(\lambda) = & 4\left[\langle\psi_{0}| \sum_{k=0}^{\min[1-s,1-q]}\sum_{p=0}^{n}\sum_{q=0}^{1}\sum_{s=0}^{1}p!q!s!k!  \binom{n}{p}\binom{n}{p} \binom{1}{q}\binom{2n-p}{q} \right.\\
    %&\left. \binom{1}{s}\binom{2n-p}{s} \binom{1-s}{k}\binom{1-q}{k} \hat{a}^{\dagger\,2n-p-q-s-k+1} \hat{a}^{2n-p-q-s-k+1}|\psi_{0}\rangle \right.\\
   % &-\left(\langle\psi_{0}|\sum_{m=1}^{\min[1-f,1-j]}\sum_{f=0}^{1} \sum_{j=0}^{1}j!f!m! \binom{1}{f}\binom{n}{f} \binom{1}{j}\binom{n}{j} \right.\left. \binom{1-j}{m}\binom{1-f}{m} f!j!m! \hat{a}^{\dagger\,n-f-j-m+1} \hat{a}^{n-f-j-m+1}|\psi_{0}\rangle\right)^2\Big]
%\end{aligned}
%\end{equation}

\begin{align}
\label{H33}
   \mathcal{F}(\lambda) = & 4\Bigg[\langle\psi_{0}| \sum_{k=0}^{\min[1-s,1-q]}\sum_{p=0}^{n}\sum_{q=0}^{1}\sum_{s=0}^{1}p!q!s!k!  \binom{n}{p}\binom{n}{p} \binom{1}{q}\binom{2n-p}{q}  \binom{1}{s}\binom{2n-p}{s} \binom{1-s}{k}\binom{1-q}{k} \nonumber \\
  & \quad \hat{a}^{\dagger\,2n-p-q-s-k+1} \hat{a}^{2n-p-q-s-k+1}|\psi_{0}\rangle  -\Bigg(\langle\psi_{0}|\sum_{m=0}^{\min[1-f,1-j]}\sum_{f=0}^{1} \sum_{j=0}^{1}f!j!m! \binom{1}{f}\binom{n}{f} \binom{1}{j}\binom{n}{j} \nonumber \\
 & \quad \binom{1-j}{m}\binom{1-f}{m}  \hat{a}^{\dagger\,n-f-j-m+1} \hat{a}^{n-f-j-m+1}|\psi_{0}\rangle\Bigg)^2\Bigg] 
\end{align}

To calculate the QFI, it is necessary to obtain the expectation values of various combinations of the operators $\hat{a}^\dagger$ and $\hat{a}$ for the probe state under consideration . To this end, we categorize the probes into two groups where the CS group comprises CSs and PACSs and the SS group consists of SSs and PASSs. For the states in the CS group, the required expectation values can be obtained by repeatedly employing the CS eigenvalue relations $\hat{a}|\alpha \rangle=\alpha|\alpha \rangle$ and $\langle\alpha|\hat{a}^{\dagger}=\alpha^{*}\langle\alpha|$ to obtain the QFI as a function of coherent amplitude $\alpha$ and phase $\phi$. In the case of the SSs and PASSs, we should obtain the correlation function $\langle \hat{a}^{\dagger\,n} \hat{a}^{m} \rangle$.
Concerning the displaced squeezed vacuum states $|\alpha,\xi\rangle = \hat{D}(\alpha)\,\hat{S}(\xi)|0\rangle$, the authors in \cite{PhysRevA.47.4474} established a formula for these correlation functions. Therefore, in the  case of SSs $(\alpha=0)$, the formula is simplified as
\begin{equation}
    \langle \hat{a}^{\dagger\,n} \hat{a}^{m} \rangle = \sum_{l=0}^{\min[n,m]}l!\binom{n}{l}\binom{m}{l}A^{l}(\frac{1}{2}B^{*})^{n-l}(\frac{1}{2}B)^{m-l}H_{n-l}(0)H_{m-l}(0)
\end{equation}
where $H_{n}(0)$ is the Hermite polynomials of order $n$ evaluated at zero, $A=\sinh^2{r}$, $B=-e^{-i\theta} \sinh{r}\cosh{r}$, respectively. 

%In this work, we compute and analyze the QFI for various probes, considering nonlinearity orders of $n=1,2,3,4$ for the specified interactions. The objective is to identify the best probe in various scenarios.
\section{Quantum metrology of nonlinearity}
\label{sec:V}
In this section, we analyze the QFI associated with the probes introduced in section \ref{sec:III} for the nonlinear generators considered in the section \ref{sec:IV}. The results are summarized in Figs. \ref{fcoh} and \ref{fsqu} while the analytical expressions of the QFI are reported in Tables \ref{tabb}, \ref{tabe} and \ref{tabl1}. 

\begin{table}
\centering
\begin{tabular}{|c|c|c|c|}
\hline
Nonlinearity order & $(\hat{a}+\hat{a}^{\dagger})^n$ & $\hat{a}^n+\hat{a}^{\dagger\,n}$ & $\hat{a}^{\dagger\,n}\hat{a}^n$ \\
\hline
$n=1$ & Eq. (A1) & $\mathcal{F}=4  $ & Eq. (B1) \\
\hline
$n=2$ & Eq. (A2) & $\mathcal{F}= (16 \alpha^2 + 8 )$  & Eq. (B2) \\
\hline
$n=3$ & Eq. (A3) & $\mathcal{F}=12 (3 \alpha^4 + 6 \alpha^2 + 2 )$ & Eq. (B3) \\
\hline
$n=4$ & Eq. (A4)& $\mathcal{F}=32 (2 \alpha^6 + 9 \alpha^4 + 12 \alpha^2 + 3)$ & Eq. (B4) \\
\hline
\end{tabular}
\caption{The QFI ($\mathcal{F}$) for the generators $(\hat{a}+\hat{a}^{\dagger})^n$, $\hat{a}^n + \hat{a}^{\dagger\,n}$ and $\hat{a}^{\dagger\,n}\hat{a}^{n}$ for the CS as probes and different nonlinearity orders $n = 1, 2, 3, 4$.}
\label{tabb}
\end{table}
\subsection{Coherent-state (CS) group}
We first consider the CS family and the corresponding QFI for CS is summarized in Table \ref{tabb} for all nonlinear interactions. We will start with the generator $(\hat{a}+\hat{a}^{\dagger})^{n}$. For the linear case $n=1$, the interaction reduces to a displacement-type transformation and QFI takes the constant value $\mathcal{F}=4$. For higher order nonlinearities $n=2,3,4$, the QFI becomes a monotonically increasing function of both coherent amplitude $\alpha$ and the nonlinear order $n$ at optimal phase $\phi=0,\pi$. We next consider the generator $\hat{a}^n+\hat{a}^{\dagger\,n}$ and $\hat{a}^{\dagger n}\hat{a}^{n}$. For $n=1$, the operator $\hat{a}+\hat{a}^{\dagger}$ again corresponds to a linear quadrature interaction with $\mathcal{F}=4$ whereas $\hat{a}^{\dagger}\hat{a}$ corresponds to the number operator and yields $\mathcal{F}=4\alpha^2$. For higher nonlinear orders, the QFI increases with both $\alpha$ and $n$, while remaining independent of phase $\phi$.

We now turn to PACSs. The analytical expressions of the QFI are reported in Table \ref{tabe}. For the interaction $(\hat{a}+\hat{a}^{\dagger})^{n}$, the behavior for $n=1$ differs qualitatively from higher nonlinearities. In this linear regime, the QFI decreases with increasing $\alpha$ and asymptotically approaches CS value $\mathcal{F}=4$ with optimal phase $\phi=\pi/2$. However, for $n=2,3,4$, the QFI increases monotonically with both the coherent amplitude and the nonlinear order, with maximal values again obtained at $\phi=0,\pi$. For the generator $\hat{a}^n+\hat{a}^{\dagger\,n}$, the QFI exhibit an oscillatory dependence on the phase through terms proportional to $\cos{2n\phi}$. The maximal QFI is obtained when $\cos{2n\phi}=-1$, which occurs at $\phi=\pi/2n$, thus the optimal phase explicitly depends on the nonlinearity order $n$. In contrast, for the generator $\hat{a}^{\dagger n}\hat{a}^{n}$, the QFI is independent of phase $\phi$ and increases monotonically both with $\alpha$ and $n$.
The left column of Fig. \ref{fcoh} summarizes these behaviors. For all nonlinear generators, PACSs exhibit larger QFI than CSs at fixed coherent amplitude.
\begin{table}
\centering
\begin{tabular}{|c|c|c|c|}
\hline
Nonlinearity order & $(\hat{a}+\hat{a}^{\dagger})^n$ & $\hat{a}^n+\hat{a}^{\dagger\,n}$ & $\hat{a}^{\dagger\,n}\hat{a}^n$ \\
\hline
$n=1$ & Eq. (C1) & \,$\mathcal{F}=\frac{4 \left(\alpha^4-2 \alpha^2 \cos 2 \phi+2 \alpha^2+3\right)}{\left(\alpha^2+1\right)^2}$ & Eq. (D1) \\
\hline
$n=2$ & Eq. (C2)  & $\mathcal{F}=\frac{8 \left(2 \alpha^6-4 \alpha^4 \cos 4 \phi+9 \alpha^4+14 \alpha^2+3\right)}{\left(\alpha^2+1\right)^2}$  & Eq. (D2)\\
\hline
$n=3$ & Eq. (C3) & $\mathcal{F}=\frac{12 \left(3 \alpha^8-6 \alpha^6 \cos 6 \phi+24 \alpha^6+59 \alpha^4+40 \alpha^2+8\right)}{\left(\alpha^2+1\right)^2}$ & Eq. (D3) \\
\hline
$n=4$  & Eq. (C4) & $\mathcal{F}=\frac{32 \left(2 \alpha^{10}-4 \alpha^8 \cos 8 \phi+25 \alpha^8+102 \alpha^6+150 \alpha^4+90 \alpha^2+15\right)}{\left(\alpha^2+1\right)^2}$  &
Eq. (D4)
\\
\hline
\end{tabular}
\caption{The QFI ($\mathcal{F}$) for the generators $(\hat{a}+\hat{a}^{\dagger})^n$, $\hat{a}^n + \hat{a}^{\dagger\,n}$ and $\hat{a}^{\dagger\,n}\hat{a}^{n}$ for PACS as probes and different nonlinearity orders $n = 1, 2, 3, 4$.}
\label{tabe}
\end{table}
\subsection{Squeezed state (SS) group}
We now focus on SS family and the analytical expression of the corresponding QFI are reported in Table \ref{tabl1}. We first investigate the interaction $(\hat{a}+\hat{a}^{\dagger})^n$. The results show that the QFI is expressed in term of the parameter, $x=\cos\theta\sinh2r+\cosh2r$, which determines the phase dependence of the squeezing contribution. The maximal QFI is obtained when $x$ reaches its maximum value, namely for $\theta=0, 2\pi$. An important feature emerges from the analytical expressions of QFI. In both SSs and PASSs, the QFI scales as powers of the parameter $x$, however, the prefactors increase much more rapidly for PASSs than for SSs as the nonlinearity order increases. The corresponding QFI ratios are
\begin{equation}
\frac{\mathcal{F}_{\mathrm{PASS}}^{1,2}}{\mathcal{F}_{\mathrm{SS}}^{1,2}} = 3, 
\qquad
\frac{\mathcal{F}_{\mathrm{PASS}}^{3}}{\mathcal{F}_{\mathrm{SS}}^{3}} = 7,
\qquad
\frac{\mathcal{F}_{\mathrm{PASS}}^{4}}{\mathcal{F}_{\mathrm{SS}}^{4}} = \frac{15}{2}.
\end{equation}
This clearly demonstrate that the enhancement produced by photon addition becomes increasingly significant as the nonlinear order increases. In particular, for higher nonlinearities, the QFI of PASSs grow substantially faster than that of SSs. This behavior indicates that photon addition strongly amplifies the higher-order quantum fluctuations and correlations that directly contribute to the variance of nonlinear generators. Consequently, the metrological advantage of PASSs becomes especially pronounced in strongly nonlinear estimation protocols, where higher-order moments play a dominant role in determining the achievable precision.

We next consider the generator $\hat{a}^n+\hat{a}^{\dagger\,n}$. It shows an increasing trend with the parameter $r$ but exhibits fluctuations in phase $\theta$. Furthermore, maximum QFI values can be achieved at multiple phase points $\theta$ based on the nonlinearity order. However, for the interaction $\hat{a}^{\dagger\,n} \hat{a}^n$, the results indicate that the QFI does not depend on the phase $\theta$, but rather increases as a function of the parameters $r$ and the nonlinearity order. For the generator $\hat{a}^n+\hat{a}^{\dagger\,n}$, the QFI ratio between PASSs and SSs exhibits different behaviors depending on the nonlinearity order. For $n=1$ and $n=2$, the ratio remains constant and equal to 3. However, for higher nonlinearities $n=3$ and $n=4$, the ratio initially increases in the small-squeezing regime and subsequently approaches a constant value for larger squeezing strengths. In contrast, for the generator $\hat{a}^{\dagger n}\hat{a}^{n}$, the behavior of the QFI ratio between PASSs and SSs differs significantly from the previous interactions. For the linear case $n=1$, the ratio remains constant and equal to 3. For higher nonlinearities, however, the ratio becomes dependent on squeezing parameter $r$. In the case of $n=2$, the ratio decreases continuously with increasing $r$. For $n=3$ and $n=4$, the relative enhancement decreases with increasing squeezing strength, especially in the experimentally relevant region i.e., in the small squeezing regime, and gradually saturates for larger $r$.  

Overall, the behavior observed for all generators show that photon addition provides the greatest metrological benefit in weak-to-intermediate squeezing regimes that are experimentally accessible with current CV quantum optical platforms. Moreover, the advantage becomes increasingly significant for strongly nonlinear estimation schemes, highlighting the potential of non-Gaussian resource engineering for enhancing precision in higher-order quantum metrology.

%We proceed with focusing on the SS group, i.e., Ss, PSSs, and PASs as probes. The results are presented in Tables \ref{tabl}, \ref{tabs}, and \ref{tabf}, respectively. At first, the interaction $(\hat{a}+\hat{a}^{\dagger})^n$ is investigated. The results indicate that the QFI increases with the nonlinearity order $n$ and squeezing parameters $r$ for this group of probes. Moreover, for the cases of PSSs and PASs, the QFI values are observed to be identical. These results are detailed as a function of the parameter $x=\cos(\theta)\sinh(2r)+\cosh(2r)$. Also, the maximum value of QFI is achieved when the parameter $x$ reaches its maximum value, which occurs at $\theta=0,\, 2\pi$.
%In the case of the generator $\hat{a}^n+\hat{a}^{\dagger\,n}$, the QFI depends on various parameters. It shows an increasing trend with the parameter $r$ but exhibits fluctuations in phase $\theta$. Furthermore, maximum QFI values can be achieved at multiple phase points $\theta$ based on the non-linearity order. However, for the interaction $\hat{a}^{\dagger\,n} \hat{a}^n$, the results indicate that the QFI does not depend on the phase $\theta$, but rather increases as a function of the parameters $r$ and the nonlinearity order.
\begin{table*}[t]
\centering
\begin{tabular}{|c|c|c|c|c|c|c|}
\hline
Nonlinearity Order
& \multicolumn{2}{c|}{$(\hat a+\hat a^\dagger)^n$}
& \multicolumn{2}{c|}{$\hat a^n+\hat a^{\dagger n}$}
& \multicolumn{2}{c|}{$\hat a^{\dagger n}\hat a^n$} \\
\cline{2-7}
& SS & PASS & SS & PASS & SS & PASS \\
\hline
$n=1$ & $\mathcal{F}=4 x$ & $\mathcal{F}=12 x$ & Eq. (E1) & Eq. (G1) & Eq. (F1) & Eq. (H1) \\
\hline
$n=2$ & $\mathcal{F}=8 x^2$ & $\mathcal{F}=24 x^2$ & Eq. (E2) & Eq. (G2) & Eq. (F2) & Eq. (H2) \\
\hline
$n=3$ & $\mathcal{F}=60 x^3$ & $\mathcal{F}=420 x^3$ & Eq. (E3) & Eq. (G3) & Eq. (F3) & Eq. (H3) \\
\hline
$n=4$ & $\mathcal{F}=384 x^4$ & $\mathcal{F}=2880 x^4$ & Eq. (E4) & Eq. (G4) & Eq. (F4) & Eq. (H4) \\
\hline
\end{tabular}
\caption{The QFI ($\mathcal{F}$) for the generators $(\hat{a}+\hat{a}^{\dagger})^n$, $\hat{a}^n + \hat{a}^{\dagger\,n}$ and $\hat{a}^{\dagger\,n}\hat{a}^{n}$ for the SS and PASS as probe and different orders of nonlinearity $n = 1, 2, 3, 4$.}
\label{tabl1}
\end{table*}

%\begin{table}
%\centering
%\begin{tabular}{|c|c|c|c|}
%\hline
%Nonlinearity order & $(\hat{a}+\hat{a}^{\dagger})^n$ & $\hat{a}^n+\hat{a}^{\dagger\,n}$ & $\hat{a}^{\dagger\,n}\hat{a}^n$ \\
%\hline
%%\hline
%$n=2$ & $\mathcal{F}=8 x^2$ & Eq. (E2)  & Eq. (F2)\\
%\hline
%$n=3$ & $\mathcal{F}=60 x^3$ & Eq. (E3) & Eq. (F3)\\
%\hline
%$n=4$ & $\mathcal{F}=384 x^4$  & Eq. (E4) & Eq. (F4) \\
%\hline
%\end{tabular}
%\caption{The QFI ($\mathcal{F}$) for the generators $(\hat{a}+\hat{a}^{\dagger})^n$, $\hat{a}^n + \hat{a}^{\dagger\,n}$ and $\hat{a}^{\dagger\,n}\hat{a}^{n}$ for the SS as probe and different orders of nonlinearity $n = 1, 2, 3, 4$.}
%\label{tabl}
%\end{table}

%\begin{table}
%\centering
%\begin{tabular}{|c|c|c|c|}
%\hline
%Nonlinearity order & $(\hat{a}+\hat{a}^{\dagger})^n$ & $\hat{a}^n+\hat{a}^{\dagger\,n}$ & $\hat{a}^{\dagger\,n}\hat{a}^n$ \\
%\hline
%$n=1$ & $\mathcal{F}=12 x$ & Eq. (G1) & Eq. (H1) \\
%\hline
%$n=2$ &$\mathcal{F}=24 x^2$& Eq. (G2)  & Eq. (H2)\\
%\hline
%$n=3$ &$\mathcal{F}=420 x^3$ & Eq. (G3) & Eq. (H3)\\
%\hline
%$n=4$ & $\mathcal{F}=2880 x^4$  & Eq. (G4) &  Eq. (H4) \\
%\hline
%\end{tabular}
%\caption{The QFI ($\mathcal{F}$) for the generators $(\hat{a}+\hat{a}^{\dagger})^n$, $\hat{a}^n + \hat{a}^{\dagger\,n}$ and $\hat{a}^{\dagger\,n}\hat{a}^{n}$ for PASS as probe and different orders of nonlinearity $n = 1, 2, 3, 4$.}
%\label{tabs}
%\end{table}

\section{optimal probes}
\label{sec:VI}
To achieve optimal performance in the estimation of the interaction parameter, it is necessary to identify the most suitable probe for various Hamiltonians from different points of view, i.e., the parameters of the probe, its energy, and also the degree of NG in the case of non-Gaussian states. Since the optimal phases have already been identified in the previous section, all comparison presented here refer to the corresponding maximum QFI values. At first, we consider the states in the CS group and then analyze the states within the SS group.
\begin{figure}
    \centering
    \begin{tabular}{ccc}
        \includegraphics[width=0.28\columnwidth]{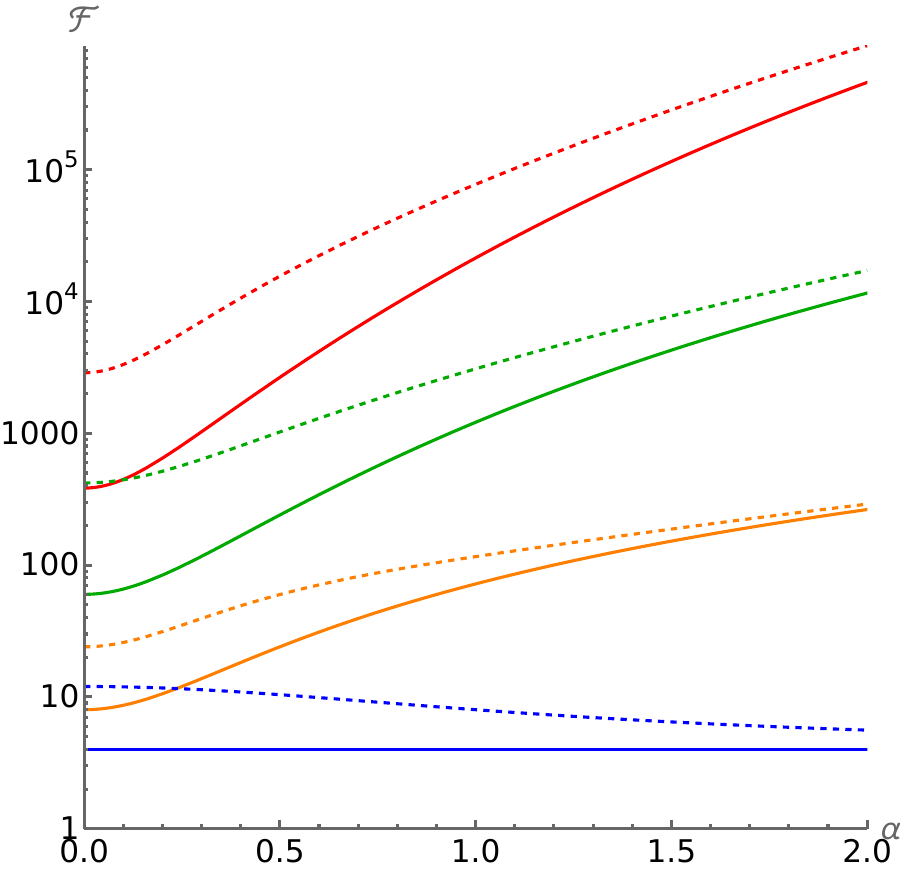} &
        \includegraphics[width=0.28\columnwidth]{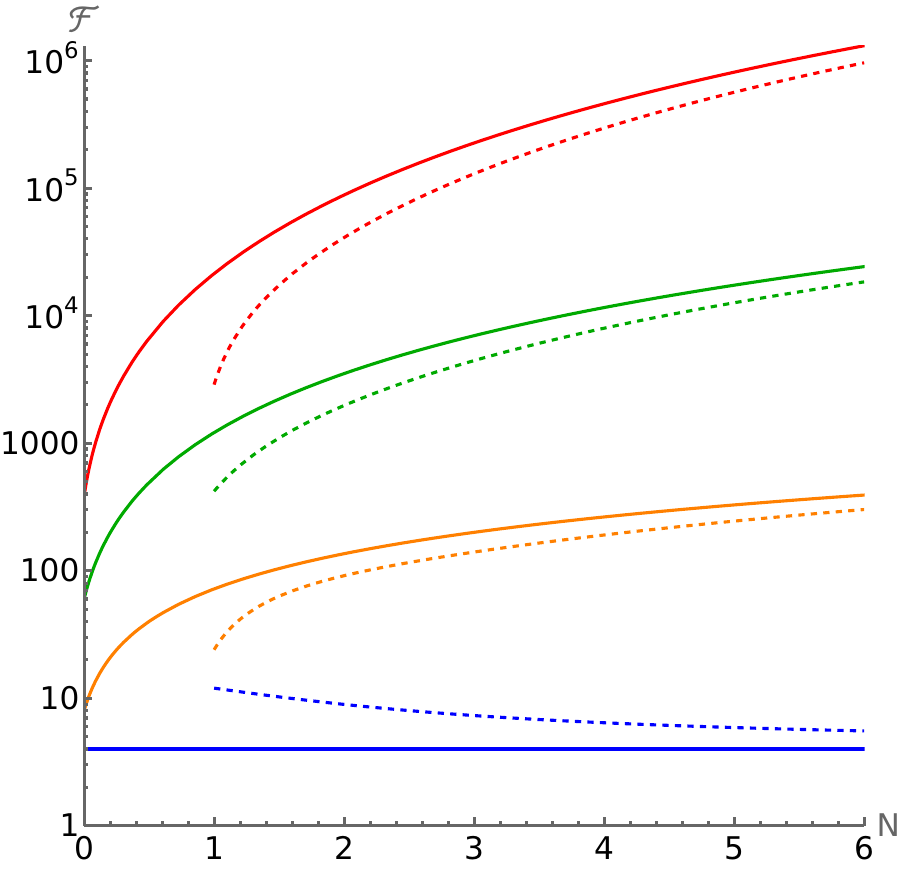} &
        \includegraphics[width=0.28\columnwidth]{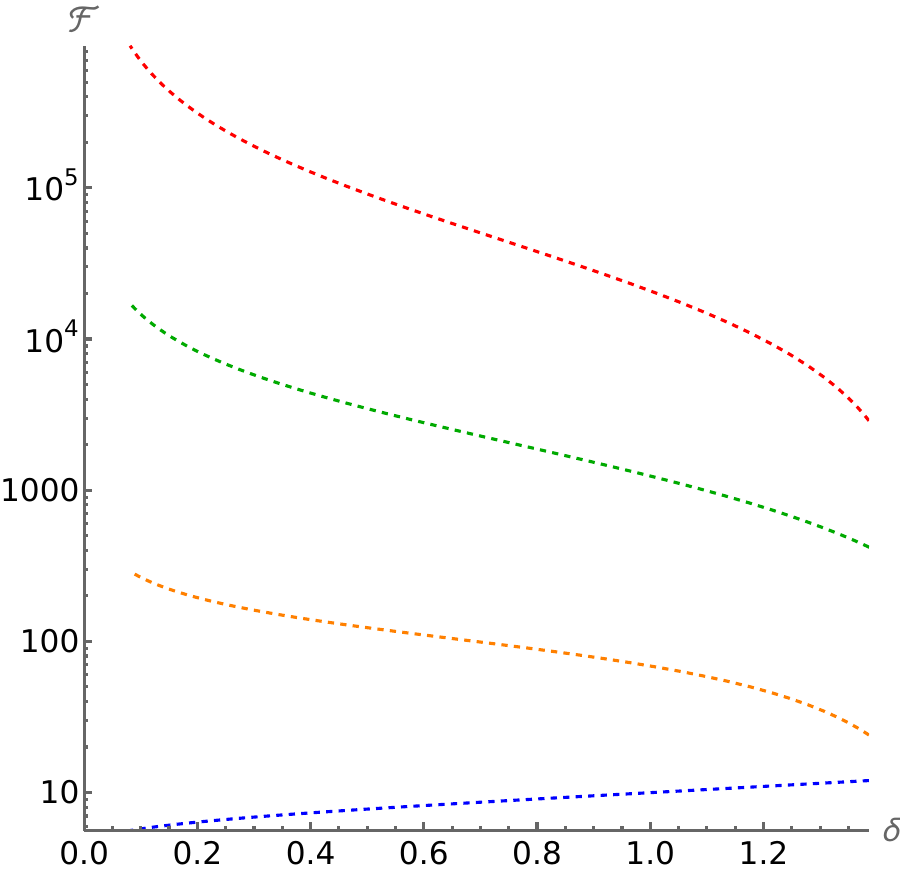} \\
        \includegraphics[width=0.28\columnwidth]{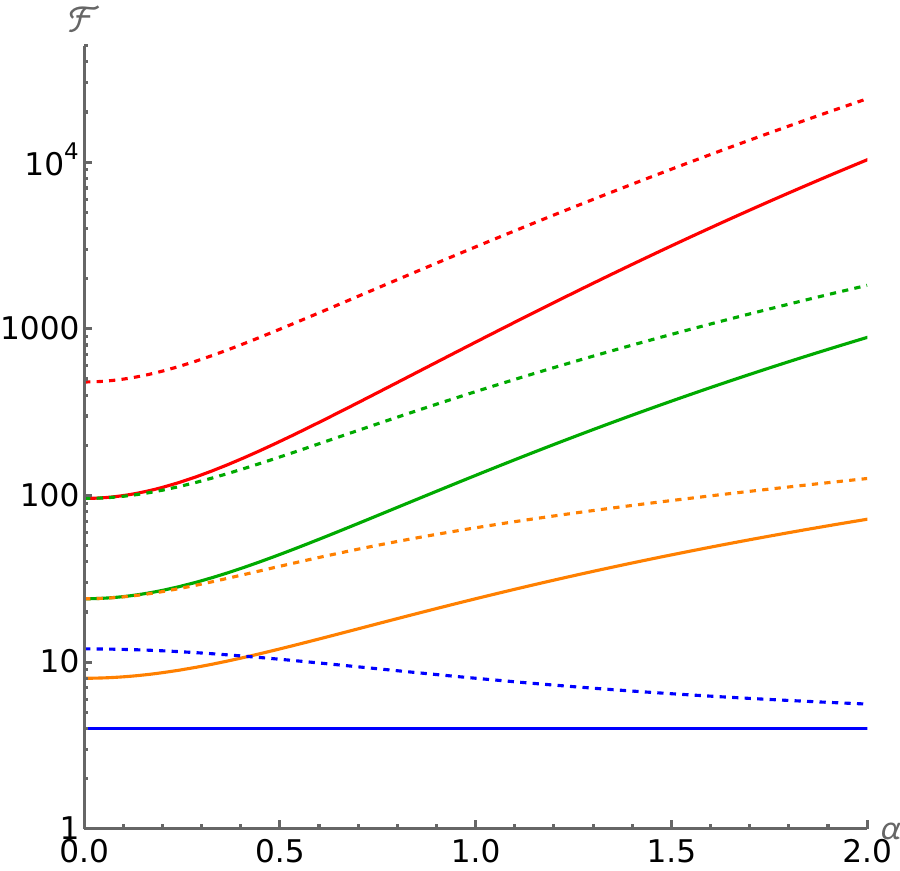} &
        \includegraphics[width=0.28\columnwidth]{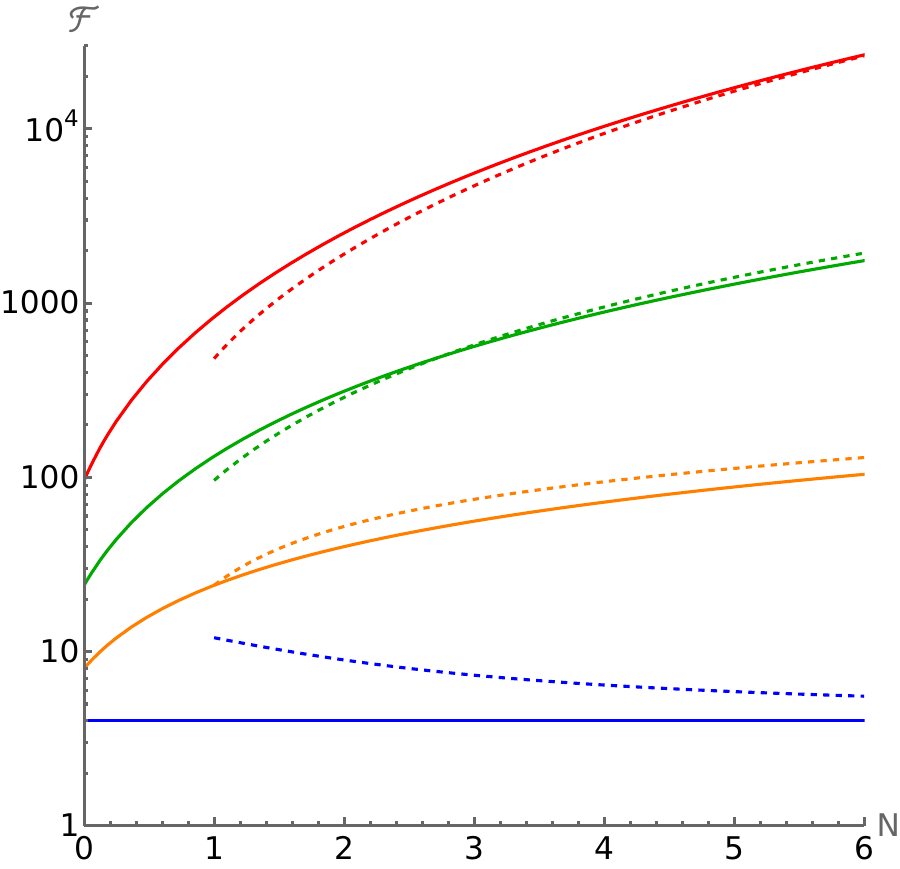} &
        \includegraphics[width=0.28\columnwidth]{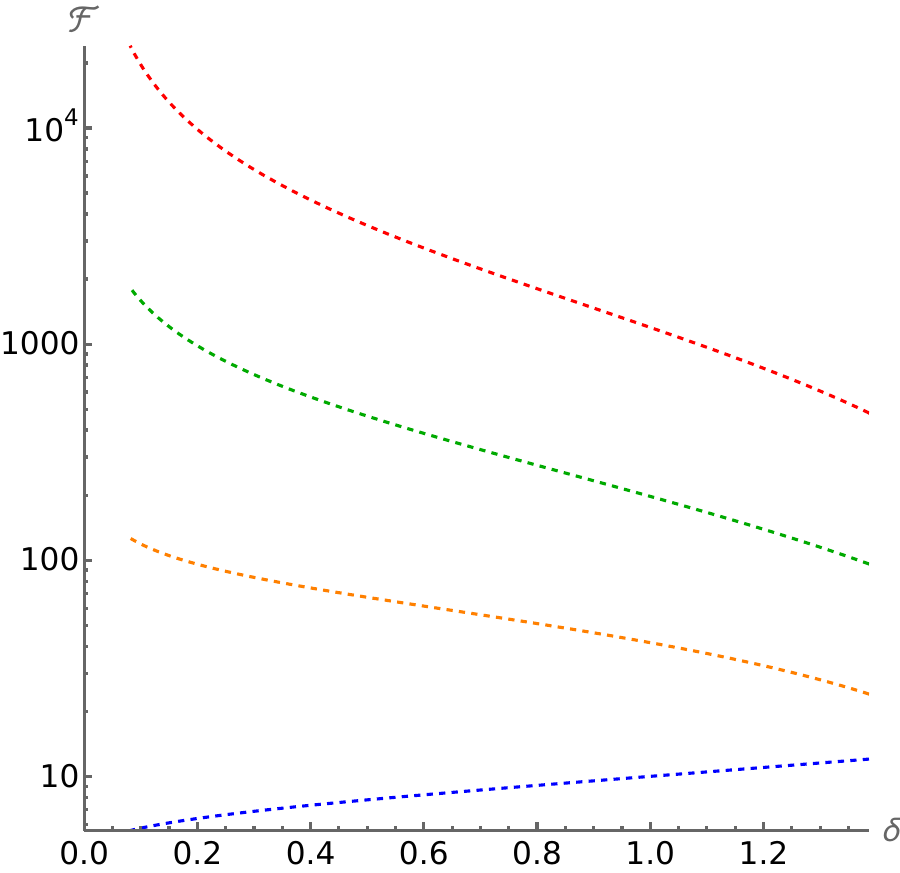} \\
        \includegraphics[width=0.28\columnwidth]{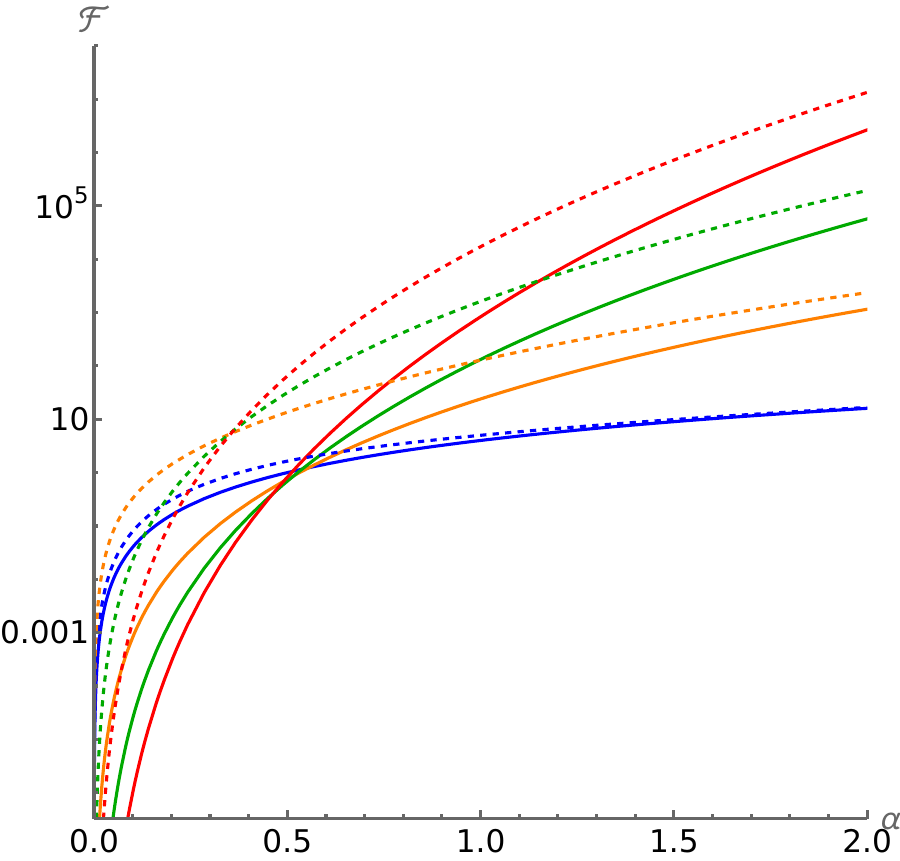} &
        \includegraphics[width=0.28\columnwidth]{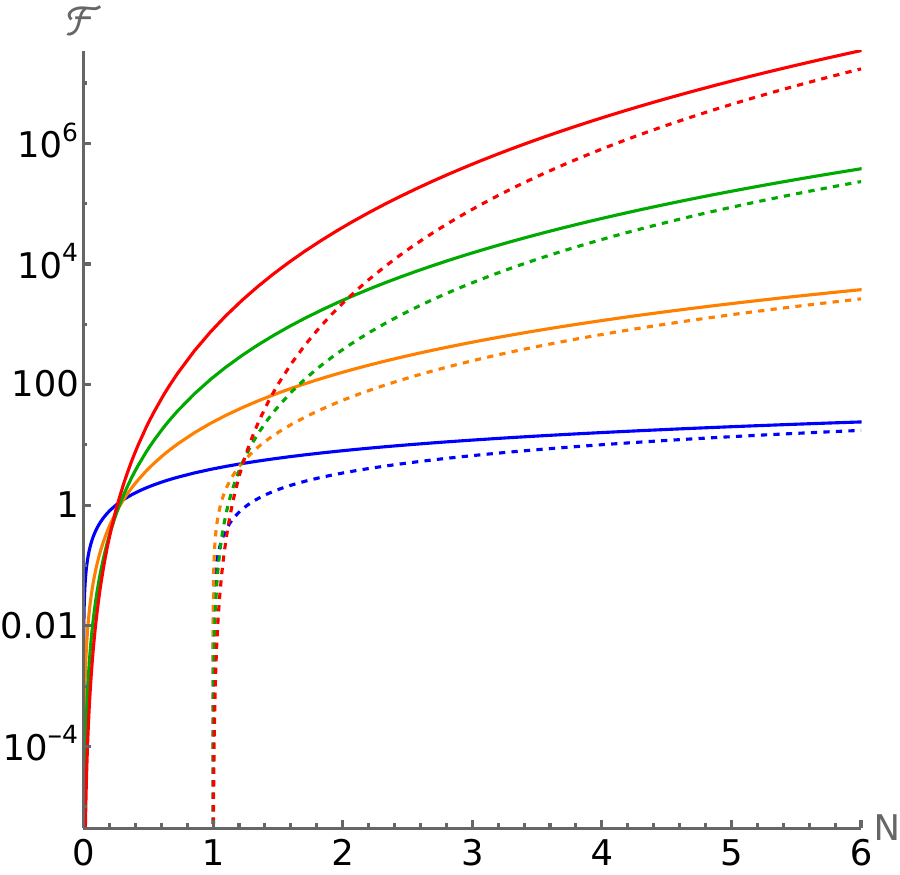} &
        \includegraphics[width=0.28\columnwidth]{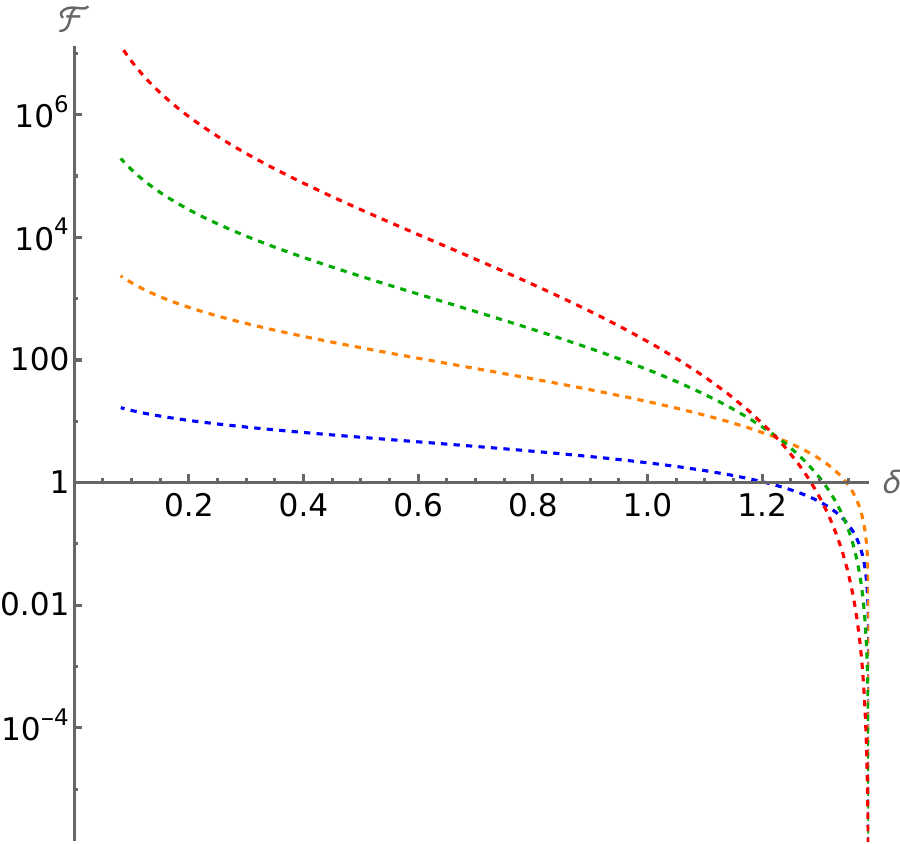} \\
    \end{tabular}
    \caption{Left column: The QFI as a function of $\alpha$. Middle column: The QFI as a function of the energy ($N$). Right column: The QFI of the PACS as a function of the degree of NG ($\delta$). The first, second, and third rows correspond to the generators $(\hat{a}+\hat{a}^{\dagger})^n$, $\hat{a}^n+\hat{a}^{\dagger n}$, and $\hat{a}^{\dagger n}\hat{a}^{n}$, respectively. Solid lines represent CSs, and dashed lines represent PACSs. The blue, orange, green and red lines are associated with the nonlinearity orders $n=1$, $n=2$, $n=3$, and $n=4$, respectively.}
    \label{fcoh}
\end{figure}
\subsection{Optimal probe in the CS group}
We start with finding the best probe based on the parameters of the probe.
%To do a fair comparison, we obtain the maximum QFI for all situations. Since the maximum value of the QFI is typically found at the boundaries phase, we set the parameters $\phi=0$ and compare the QFI of different probes. The only case where this rule does not apply is the interaction $\hat{a}^n+\hat{a}^{\dagger\,n}$ and for the PACSs. In this case, the maximum value cannot be achieved when $\phi=0$. Therefore, we consider the first value of $\phi$ where the QFI reaches its maximum. 
The left column of Fig.\ref{fcoh} shows the QFI as a function of CS amplitude $\alpha$ for CSs and PACSs. For all generators, the QFI increases with increasing $\alpha$ and the growth becomes progressively stronger as the nonlinear order increases. This behavior indicates that nonlinear generators strongly amplify the contribution of higher-order photon number fluctuations to the estimation precision. The absolute difference between the curves decreases with increasing $\alpha$ and asymptotically approaches CS limit.
%especially for the kerr-type interaction $\hat a^{\dagger n}\hat a^n$, where the QFI grows by several orders of magnitude even for moderate values of $\alpha$.This behavior indicates that nonlinear generators strongly amplify the contribution of higher-order photon number fluctuations to the estimation precision. 
An important exception appears for the $(\hat{a}+\hat{a}^{\dagger})^n$ and $\hat{a}^n+\hat{a}^{\dagger n}$ in linear case $n=1$. For these generators, the CS QFI is independent of $\alpha$, whereas the PACS QFI initially decreases with increasing $\alpha$ and asymptotically approaches the CS limit. An important observation concerns the comparison between CSs and PACSs. At fixed $\alpha$, PACSs provide larger QFI than CSs, and the advantage becomes evident for the nonlinear order $(n>1)$. 
%In particular, for $n=3$ and $n=4$, the separation between the two curves grows rapidly with $\alpha$, showing that photon addition significantly enhances the sensitivity of nonlinear transformations. 
%This enhancement originates from the stronger higher-order moments introduced by photon addition, which directly contribute to the variance of nonlinear generators.

%We present the QFI in terms of $\alpha$ in the left column of Fig. \ref{fcoh} for different interactions and the nonlinearity order $n=1,2,3,4$. Since Cs and PSCs exhibit similar behavior, we will only present the results for Cs and PACs. It is observed that for all interactions, as the nonlinearity and $\alpha$ values increase, the QFI also increases, resulting in enhanced performance for both states. A comparison between the two states indicates that PACs exhibit superior performance compared to Cs. 

Now, we analyze this topic from the energy point of view. This can be achieved by examining the behavior of the maximal QFI for various probe states at the same initial energy. 
The total number $N$ (mean number of photons) of an arbitrary state $| \zeta \rangle$, is obtained by $N= \langle 
\zeta | a^{\dagger} a 
|\zeta \rangle$. Therefore, considering the states mentioned above, we have
\begin{equation}
N_{CS}=|\alpha|^2,\,\,\,\,\,\,N_{PACS}=\frac{1+3 |\alpha|^2+|\alpha|^4}{1+|\alpha|^2},
\end{equation}
The QFI as a function of the $N$ is illustrated in the middle column of Fig. \ref{fcoh}. As we observe, an increase in the parameter $\alpha$ leads to a corresponding rise in energy, thereby improving the performance of both states. Moreover, a comparison of the two states reveals that CSs demonstrate better performances in comparison to PACSs except for the linear case of $(\hat{a}+\hat{a}^{\dagger})^n$ and the linear and quadratic cases of $\hat{a}^n+\hat{a}^{\dagger n}$.

Let us now investigate how the metrological performance is related to the NG of the probe. 
A convenient measure of NG is provided by the relative entropy of non-gaussianity \cite{PhysRevA.78.060303,PhysRevA.82.052341}, which is defined as 
\begin{equation}\label{nG1}
\delta[\varrho]=S(\rho||\rho_{G})=S(\rho_{G})-S(\rho)
\end{equation}
where $S(\rho)$ denotes the von-Neumann entropy of a state $\rho$, and 
$\rho_{G}$ is a Gaussian reference state with the same first and second moments of 
the non-Gaussian state under investigation. The von-Neumann entropy of a single-mode Gaussian state, $S(\rho_{G})$, with covariance matrix $\sigma$, can be written as
\begin{equation}\label{von}
    S(\rho_{G})=h(x)=(x+\frac{1}{2}) \ln (x+\frac{1}{2})-(x-\frac{1}{2}) \ln (x-\frac{1}{2})
\end{equation}
where $x=\sqrt{\det(\sigma)}$.
Since we are dealing with pure non-Gaussian states, $S(\rho)=0$ and the 
NG measure may be written as:
\begin{equation}\label{ng3}
\delta[\varrho]=h(\sqrt{\det\sigma})
\end{equation}
For PACS, non-Gaussianity as a function of $\alpha$ is given by
\begin{equation}\label{ngco}
%\begin{split}
    \delta(\alpha) = h\left(\frac{1}{2}\sqrt{\frac{8}{\left(\alpha^2+1\right)^3}+1}\right)\\
    %
    %& \left(\frac{1}{2} \sqrt{\frac{8}{\left(\alpha^2+1\right)^3}+1}+\frac{1}{2}\right)  \log \left(\frac{1}{2} \sqrt{\frac{8}{\left(\alpha^2+1\right)^3}+1}+\frac{1}{2}\right) \\
    %& -\left(\frac{1}{2} \sqrt{\frac{8}{\left(\alpha^2+1\right)^3}+1}-\frac{1}{2}\right) \log \left(\frac{1}{2} \sqrt{\frac{8}{\left(\alpha^2+1\right)^3}+1}-\frac{1}{2}\right)
%\end{split}
\end{equation}
The QFI of PACSs as a function of NG for different interactions is shown in the right column of Fig. \ref{fcoh}. The behavior for $n=1$ again differs from higher nonlinearities. In the linear regime, the QFI changes only weakly with $\delta$, indicating that NG plays a negligible role in displacement estimation. However, for $n=2,3,4$, the QFI decreases monotonically as the NG increases. Thus, PACSs with lower NG provide better estimation than highly non-Gaussian states. Moreover, comparison between the middle and right columns shows that lower NG corresponds to states with larger energy. This establishes an inverse relation between NG and metrological performance within the CSs family.
Overall, Fig. \ref{fcoh} reveals a clear distinction between the linear and nonlinear regimes. For the linear case corresponding to displacement estimation, photon addition provides metrological advantage. In contrast, for other cases, although PACSs exhibit enhanced QFI at fixed $\alpha$, the energy based comparison demonstrates that this enhancement is not a genuine non-Gaussian advantage. The improvement mainly originates from the additional energy carried by photon-added states. Consequently, within the CSs family, CSs remain the optimal practical probes from the resource point of view, since the same or better estimation precision can always be achieved using Gaussian CS with sufficiently large energy, without requiring de-Gaussification procedures.

%It is evident that as $\delta$ increases, the QFI decreases except for $n=1$ of the generators $(\hat{a}+\hat{a}^{\dagger})^n$ and $\hat{a}^n+\hat{a}^{\dagger n}$. Therefore, a PACs with lower NG performs better than one with high NG. Moreover, comparing the middle and right columns shows that a lower NG signifies a state with greater energy.
\begin{figure}
    \centering
    \begin{tabular}{c@{\hspace{5mm}}c@{\hspace{5mm}}c}
\includegraphics[width=0.3\columnwidth]{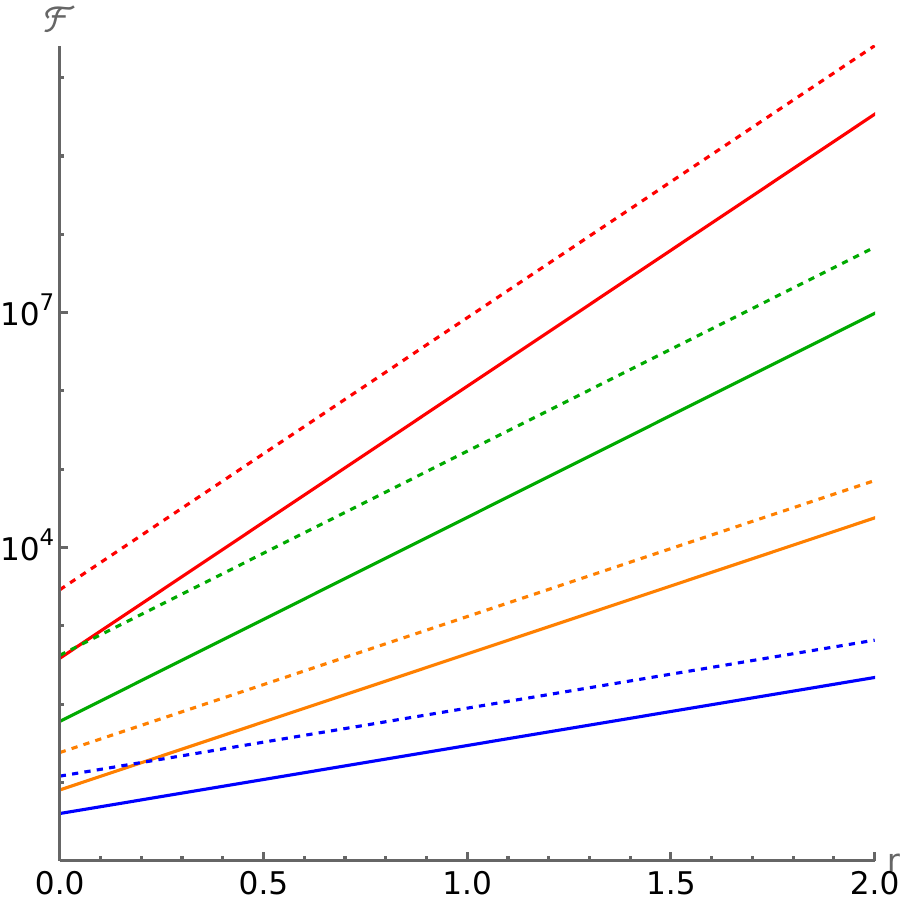}&
\includegraphics[width=0.3\columnwidth]{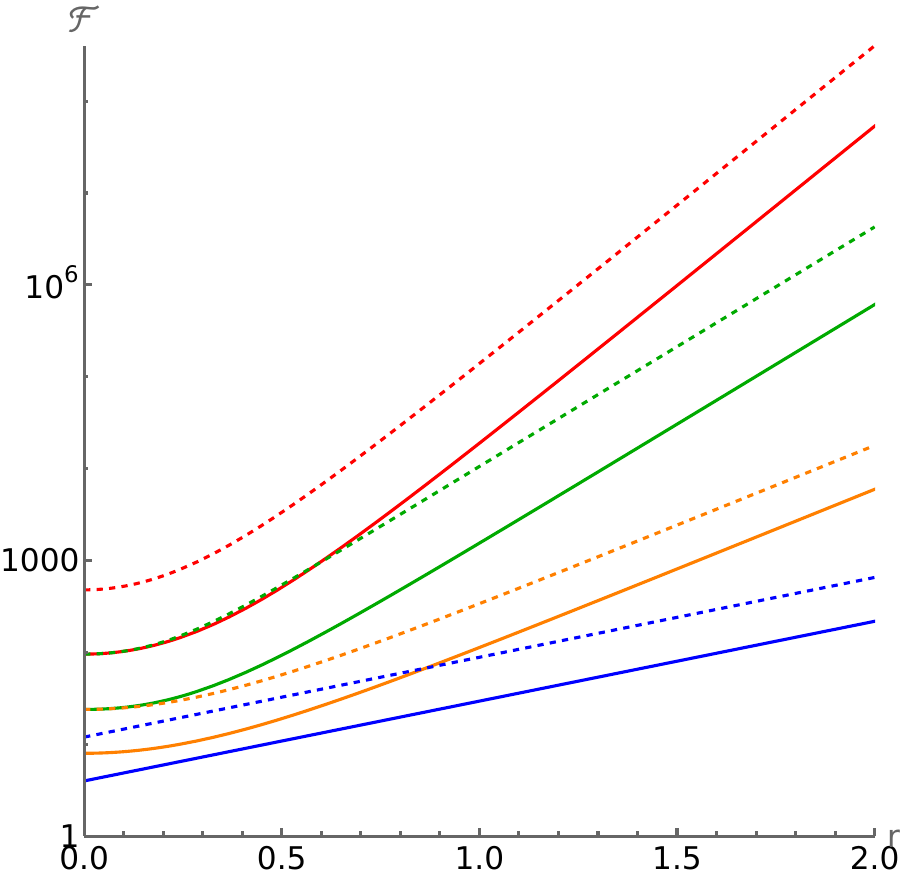}&
\includegraphics[width=0.32\columnwidth]{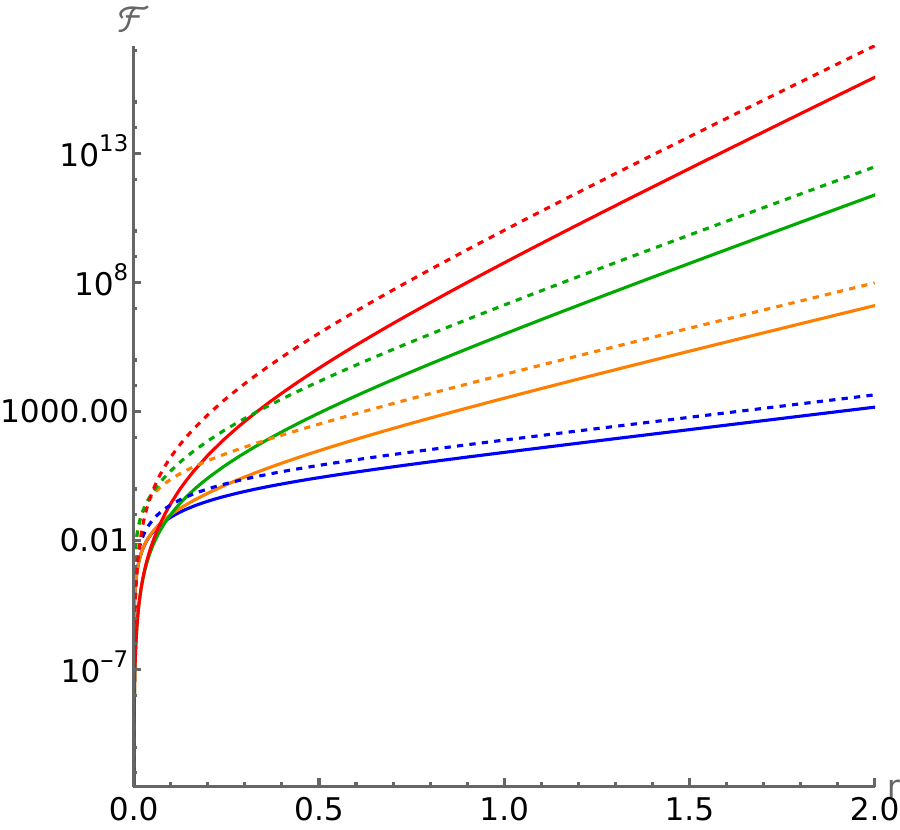}\\
\includegraphics[width=0.3\columnwidth]{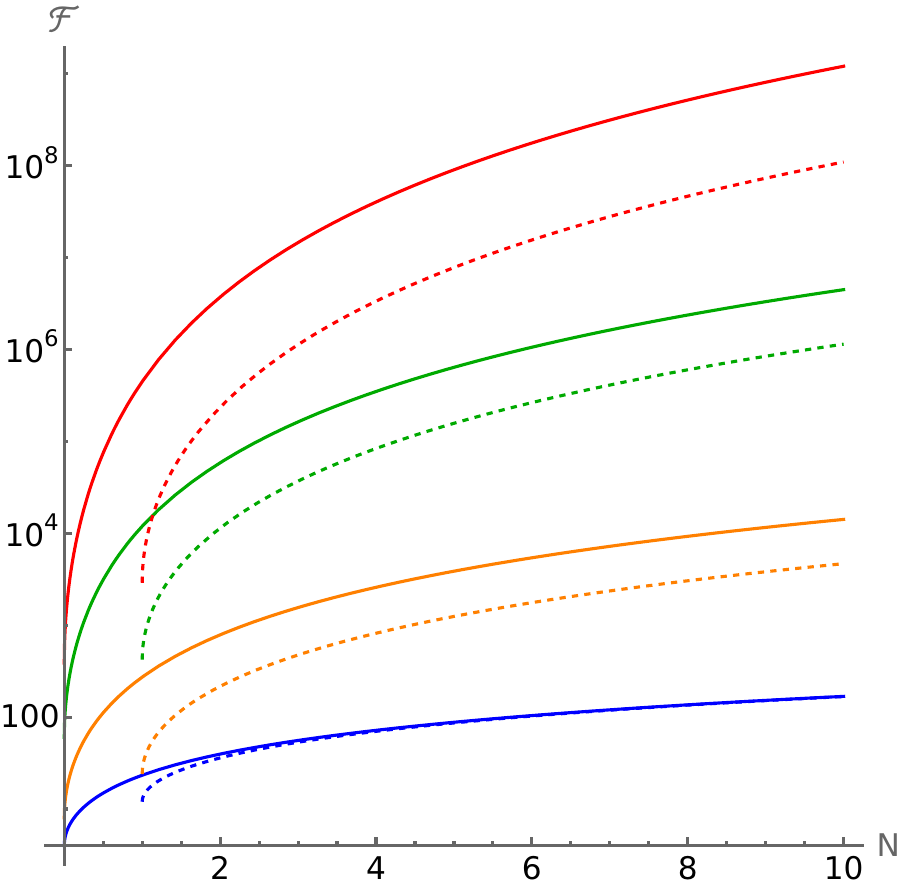}&
\includegraphics[width=0.3\columnwidth]{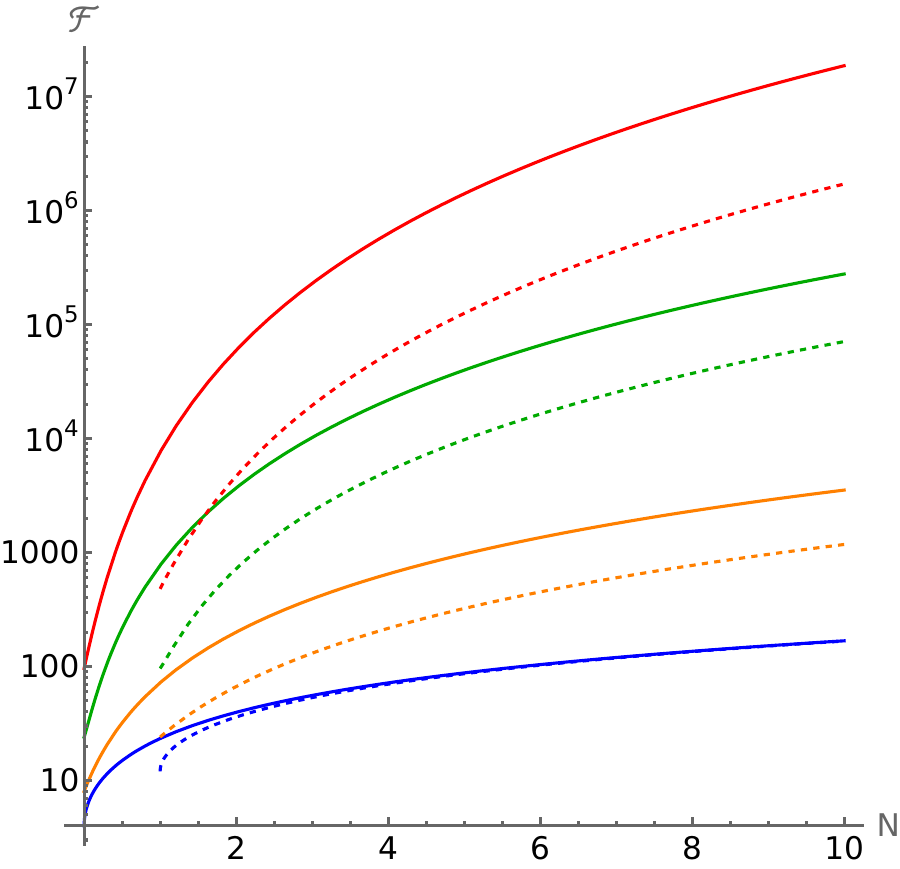}&
\includegraphics[width=0.3\columnwidth]{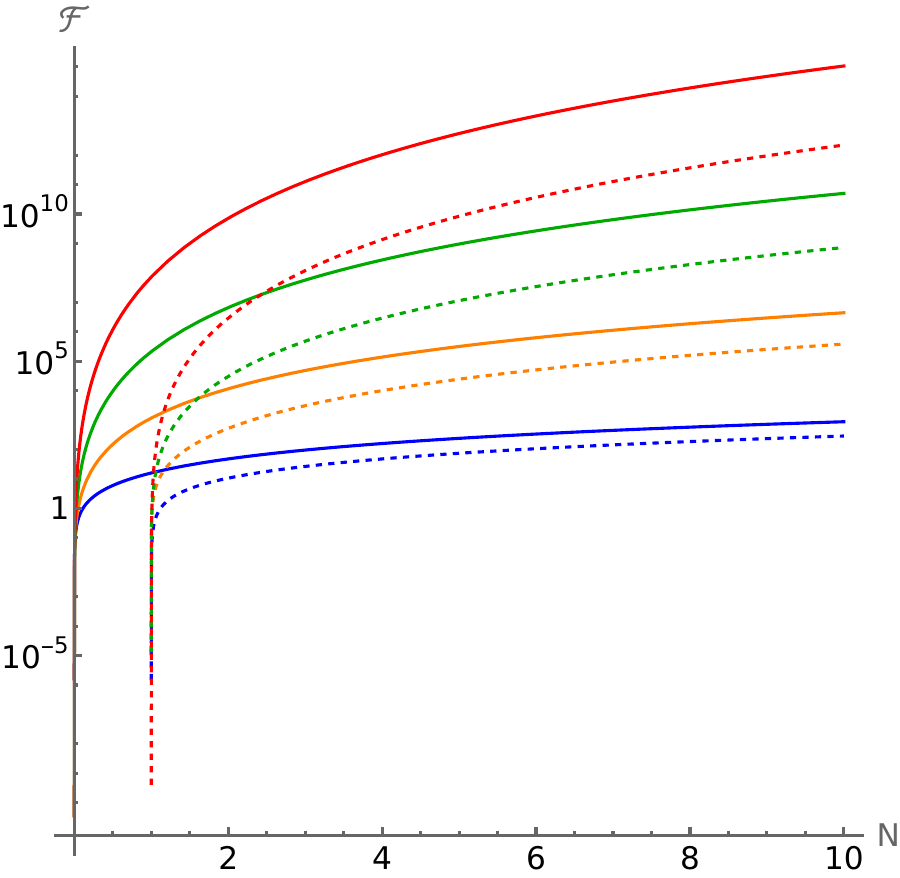}\\
\end{tabular}
\caption{First line: The QFI as a function of $r$. Second line: The QFI as a function of the energy ($N$). The first, second, and third columns correspond to the generators $(\hat{a}+\hat{a}^{\dagger})^n$, $\hat{a}^n+\hat{a}^{\dagger n}$, and $\hat{a}^{\dagger n}\hat{a}^n$, respectively. Solid lines represent SSs, dashed lines represent PASSs. The blue, orange, green and red lines are associated with the nonlinearity orders $n=1$, $n=2$, $n=3$, and $n=4$, respectively.}
\label{fsqu}
\end{figure}
\subsection{Optimal probe in the SS group}
The first line of Fig.\ref{fsqu} shows the QFI in terms of the squeezing parameter $r$ for SSs and PASSs. The result indicate that the QFI increases rapidly with both squeezing parameter $r$ and the nonlinearity order for all generators. This enhancement becomes particularly more pronounced for higher order nonlinearities, especially for Kerr-type interaction, where the QFI grows by several orders of magnitude even within moderate squeezing regime. A clear hierarchy among the probes also emerges. For all generators, PASSs exhibit systematically larger QFI than standard SSs when compared at the same squeezing parameter.  

%The results indicate that for the generator $(\hat{a}+\hat{a}^{\dagger})^n$, PASs and PSSs exhibit the same behavior. Therefore, we only present the results for Ss and PASs. Furthermore, after examining the operators $\hat{a}^n + \hat{a}^{\dagger\,n}$ and $\hat{a}^{\dagger\,n}\hat{a}^n$, it is observed that PASs and PSSs demonstrate very similar behaviors, to the extent that they are almost identical in certain values of $r$. Nevertheless, the results indicate that the PASs and PSCs exhibit superior performance in all generators compared to the Ss.

We next proceed by assessing probes in terms of energy. For the state of this group, the energy is given by
\begin{equation}
N_{SS}=|\nu|^2,\,\,\,\,\,\,N_{PASS}=1+3|\nu|^2,
\end{equation}

We illustrate the QFI as a function of the energy in the second line of Fig. \ref{fsqu}. In contrast to parameter based comparison, the fixed energy analysis shows that the SSs provide larger QFI than PASSs for the same photon number budget. This behavior can be understood physically from the different ways in which the available energy is distributed within the probes. In SSs, the entire energy resource contributes coherently to squeezing correlations, on the other hand, in PASSs, part of energy is consumed in creating the non-Gaussian excitation associated with photon addition. Although these operations enhance higher-order fluctuations, they do not use the available energy as efficiently as pure squeezing. Consequently, at fixed energy, SSs achieve larger QFI.

Finally, we  discuss the role of NG itself. Standard SSs are Gaussian and therefore NG is vanishing $\delta=0$, whereas the PASSs posses a finite and constant NG.
\begin{equation}
    \delta_{PASS}=2\log2
\end{equation}
An important feature is that this NG is intrinsic to PASSs and does not depend on squeezing strength $r$. Thus, unlike ordinary SSs, these probes retain a persistent non-Gaussian structure throughout the entire squeezing regime. As a consequence, the increase in squeezing acts on a state that already contains additional quantum features beyond Gaussian fluctuations, makes the probe more responsive to parameter variations generated by nonlinear interactions. In this sense, the fixed NG introduced by photon addition or subtraction acts as a catalyst that allows the squeezing resource to be converted into metrological sensitivity more effectively, especially for higher-order nonlinear interactions. Although SSs provide the highest QFI at fixed energy, increasing their QFI requires progressively larger squeezing strengths. By contrast, comparison of the plots in the two lines of Fig. \ref{fsqu} shows that similar QFI values can be obtained with PASSs at substantially lower squeezing parameters. Therefore, photon addition becomes particularly beneficial when applied to an already nonclassical state, such as SS, reducing the squeezing required to achieve a given estimation precision. This makes PASSs a promising alternative for nonlinear quantum metrology in experimentally accessible weak-to-intermediate squeezing regimes. 
%%%
\section{Conclusions}
\label{sec:VII}
In this work, we have studied the  estimation of the nonlinear coupling strengths of nonlinear media. As probes, we have considered Gaussian states (coherent and squeezed vacuum) together with their non-Gaussian counterparts obtained by photon addition and subtraction. Our goal was to identify which probe performs best for three different nonlinear Hamiltonians which are of interest in CV quantum technology, specifically those of the form \((\hat{a}+\hat{a}^\dagger)^n\), \(\hat{a}^n+\hat{a}^{\dagger n}\), and \(\hat{a}^{\dagger n}\hat{a}^n\), with \(n=1,2,3,4\). We compared the probes from three complementary perspectives: their intrinsic parameters, their energy, and, for non-Gaussian states, their degree of non-Gaussianity.
Our results indicate the de-Gaussification by photon subtraction and addition enhances metrological power when the initial state  possesses nonclassical features, such as squeezing. 

For coherent probes, photon addition yield a larger quantum Fisher information (QFI) than standard coherent-states when compared at the same coherent amplitude. However, this advantage disappears when the comparison is made at fixed energy. In fact, any precision achievable with a PACS can also be obtained with a  
coherent-state simply by increasing its energy. Moreover, within the PACS family, states with a lower degree of non-Gaussianity perform better, and these in turn correspond to states with higher energy. Therefore, photon addition onto coherent-states does not provide a genuine metrological resource. Rather, it merely redistributes energy in a less efficient way.

The situation is radically different for squeezed vacuum states. Photon-added and photon-subtracted squeezed states (PASSs and PSSSs) significantly outperform standard squeezed states when compared at the same squeezing parameter. More importantly, this advantage persists, and often grows, when the comparison is carried out at equal energy. Reaching the same QFI level with a Gaussian squeezed state would require impractically large squeezing strengths, which are challenging to realize experimentally. The advantage offered by PASSs is particularly significant for higher-order nonlinearities and in the weak-to-intermediate squeezing regime, which is accessible in current experimental platforms. For the Kerr-type interaction and for higher-order quadrature nonlinearities, the QFI enhancement grows rapidly with nonlinear order, making de-Gaussified squeezed states the probes of choice for estimating nonlinear couplings. In contrast to the coherent case, here a higher degree of non-Gaussianity correlates with better performance and higher energy. Thus, de-Gaussification is genuinely beneficial when applied to already nonclassical states.

Taken together, our findings suggest a clear hierarchy: photon addition and subtraction are effective second-level resources. They enhance metrological power when the initial state  possesses nonclassical features, such as squeezing. When applied to quasi-classical states like coherent-states, they offer no fundamental advantage over simply increasing the signal energy. From a practical perspective, our results indicate that for estimating nonlinear couplings in CV systems, non-Gaussian probes derived from squeezed vacuum are promising, especially in the weak-to-intermediate squeezing regime accessible with current technology. In situations where squeezing is not available, coherent probes remain the most efficient choice.

\acknowledgements
MGAP and M. acknowledge support by EU and MUR under the project PRIN22 G53D23001110006-RISQUE. M. acknowledges partial support from the Qmet Tech Foundation under Qmet PARIMANA Post-Doctoral Fellowship program 2025, funded by the Department of Science and Technology (DST), GoI, under the National Quantum Mission (Sanction Order No. Qmet/2025-10/HRD/PARIMANA/SL/PDF/QMET-515752). PvL acknowledges support from the EU/BMFTR via CLUSTEC and via ShoQC (Quant-ERA), and from the BMFTR (former BMBF) in Germany through QuKuK and QuaPhySI.
\appendix
\section{The QFI of the interaction of $(\hat{a}+\hat{a}^{\dagger})^n$ for CS as probe and different orders of nonlinearity $n = 1, 2, 3, 4$.}
\setcounter{equation}{0}
\renewcommand{\theequation}{A\arabic{equation}}

\begin{align}
     \mathcal{F} &= 4 \\
    \mathcal{F} &= 8 \left(4 \alpha^2 \cos 2 \phi+4 \alpha^2+1\right) \\
    \mathcal{F} &= 12 \left(6 \alpha^4 \cos 4 \phi+18 \alpha^4+24 \alpha^2+24 \left(\alpha^4+\alpha^2\right) \cos 2\phi +5\right) \\
    \mathcal{F} &= 64 \left(2 \alpha^6 \cos 6 \phi+3\left(7 \alpha^4 +4\alpha^6 \right) \cos 4 \phi+6 \left( 8 \alpha^2+14 \alpha^4 +5 \alpha^6 \right) \cos 2 \phi +20 \alpha^6 +63 \alpha^4 +48 \alpha^2 +6\right)
\end{align}

\section{The QFI of the interaction of $\hat{a}^{\dagger\,n}\hat{a}^{n}$ for CS as probe and different orders of nonlinearity $n = 1, 2, 3, 4$.}

\setcounter{equation}{0}
\renewcommand{\theequation}{B\arabic{equation}}

\begin{align}
     \mathcal{F} &= 4 \alpha^2\\
    \mathcal{F} &= 8 (2 \alpha^6 + \alpha^4) \\
    \mathcal{F} &= 4 (9 \alpha^{10} + 18 \alpha^8 + 6 \alpha^6) \\
    \mathcal{F} &= 4 (16 \alpha^{14} + 72 \alpha^{12} + 96 \alpha^{10} + 24 \alpha^8)
\end{align}

\section{The QFI of the interaction $(\hat{a}+\hat{a}^{\dagger})^n$ for PACS as probe and different orders of nonlinearity $n = 1, 2, 3, 4$.}
\setcounter{equation}{0}
\renewcommand{\theequation}{C\arabic{equation}}

\begin{align}
\mathcal{F} &= \frac{4 \left(\alpha^4-2 \alpha^2 \cos 2 \phi +2 \alpha^2+3\right)}{\left(\alpha^2+1\right)^2} \\
\mathcal{F} &= \frac{8 \left(4 \alpha^6-4 \alpha^4 \cos 4 \phi +13 \alpha^4+18 \alpha^2+4 \left(\alpha^4+2 \alpha^2+3\right) \alpha^2 \cos 2 \phi +3\right)}{\left(\alpha^2+1\right)^2}
\end{align}

\begin{equation}
\begin{split}
\mathcal{F} = \frac{12 \left(-6 \alpha^6 \cos 6 \phi +215 \alpha^4+154 \alpha^2\right)}{\left(\alpha^2+1\right)^2}+ \frac{12\left(18 \left(\alpha^2+6\right) \alpha^6+6 \left(\alpha^2+1\right)^2 \alpha^4 \cos 4 \phi\right)}{\left(\alpha^2+1\right)^2} \\
\quad + \frac{12\left(6 \left(\alpha^2+1\right) \left(4 \alpha^4+17 \alpha^2+19\right) \alpha^2 \cos 2 \phi +35\right)}{\left(\alpha^2+1\right)^2}
\end{split}
\end{equation}

\begin{equation}
\begin{split}
\mathcal{F} =\frac{64 \left(20 \alpha^{10}-2 \alpha^8 \cos 8 \phi+193 \alpha^8+618 \alpha^6+801 \alpha^4+420 \alpha^2 +45 \right)}{\left(\alpha^2+1\right)^2} \\
\quad + \frac{64\left(2 \left(\alpha^4+2 \alpha^2-3\right) \alpha^6 \cos 6 \phi +\left(12 \alpha^6+85 \alpha^4+174 \alpha^2+105\right) \alpha^4 \cos 4 \phi \right)}{\left(\alpha^2+1\right)^2}\\
\quad + \frac{128 \alpha^2 \left(15 \alpha^8 + 136 \alpha^6 +399 \alpha^4 +450 \alpha^2 +180\right) \cos 2 \phi}{\left(\alpha^2+1\right)^2} 
\end{split}
\end{equation}

\section{The QFI for the generators $\hat{a}^{\dagger\,n} \hat{a}^{n}$ for PACS as probe and different orders of nonlinearity $n = 1, 2, 3, 4$.}
\setcounter{equation}{0}
\renewcommand{\theequation}{D\arabic{equation}}

\begin{align}
    \mathcal{F} &= \frac{4 \alpha^2 \left(2+2 \alpha^2+\alpha^4\right)}{\left(\alpha^2+1\right)^2} \\
    \mathcal{F} &= \frac{8 \alpha^2 \left(2 \alpha^6+11 \alpha^4+15 \alpha^2+4\right)}{\left(\alpha^2+1\right)}\\
    \mathcal{F} &=\frac{12 \alpha^4 \left(3 \alpha^{10}+36 \alpha^8+140 \alpha^6+214 \alpha^4+128 \alpha^2+18\right)}{\left(\alpha^2+1\right)^2}\\
    \mathcal{F} &= \frac{32 \alpha^6\left(2 \alpha^{12}+37 \alpha^{10} +238 \alpha^8+654 \alpha^6+786 \alpha^4+375 \alpha^2+48\right)}{\left(\alpha^2+1\right)^2}
\end{align}

\section{The QFI, $\mathcal{F}$, for the generators $\hat{a}^n + \hat{a}^{\dagger\,n}$ for SS as probe and different orders of nonlinearity $n = 1, 2, 3, 4$.}
\setcounter{equation}{0}
\renewcommand{\theequation}{E\arabic{equation}}

\begin{align}
\mathcal{F} &= 4 \cos \theta  \sinh 2 r+4\cosh 2 r \\
\mathcal{F} &= 5+3 \cosh 4r +\left(-2+4 \cos 2 \theta \right) \sinh^{2}2 r \\
\mathcal{F} &=\frac{3}{4} \left( 27 \cosh 2r +5\left( \cosh 6r + 4 \cos 3\theta \sinh^{3} 2r\right)\right) \\
\mathcal{F} &= 6 \left(5 +10 \cosh 4r + \cosh 8r + 8 \cos 4\theta \sinh^{4} 2r\right)
\end{align}

\section{The QFI ($\mathcal{F}$) for the generator $\hat{a}^{\dagger\,n}\hat{a}^{n}$ for SS  as probe and different nonlinearity orders $n = 1, 2, 3, 4$.}
\setcounter{equation}{0}
\renewcommand{\theequation}{F\arabic{equation}}

\begin{align}
\mathcal{F} &= \cosh 4r -1 \\
\mathcal{F} &= 2 \sinh^{2}2r  (-12 \cosh 2 r+6 \cosh 4r +7) \\
\mathcal{F} &=\frac{9}{4} \sinh^{4}r \cosh^{2}r (923 \cosh 2r -670 \cosh 4 r+565 \cosh 6r -434) \\
\mathcal{F} &=9 \sinh^{4}r \cosh^{2}r (5967 \cosh 2 r -5510 \cosh 4 r +4875 \cosh 6r -3920 \cosh 8 r +1750 \cosh 10 r-3066)
\end{align}

\section{The QFI ($\mathcal{F}$) for the generator $\hat{a}^n + \hat{a}^{\dagger\,n}$  for PASS as probe and different orders of nonlinearity $n = 1, 2, 3, 4$.}
\setcounter{equation}{0}
\renewcommand{\theequation}{G\arabic{equation}}
\begin{align}
\mathcal{F} &= 12 \left( \cosh 2 r+ \cos \theta \sinh 2 r \right) \\
\mathcal{F} &= 6\left(3+\cosh 4r +2\cos 2\theta \sinh^{2} 2r\right) \\
\mathcal{F} &= \frac{3}{4} \left(93 \cosh 2r + 35\left( \cosh 6r + 4 \cos 3\theta \sinh^{3} 2r\right)\right) \\
\mathcal{F} &= 15 \left(5 + 24 \cosh 4r +3 \cosh 8r + 24 \cos 4\theta \sinh^{4} 2r\right)
\end{align}

\section{The QFI ($\mathcal{F}$) for the generator $\hat{a}^{\dagger\,n} \hat{a}^n  $  for PASS as probe and different orders of nonlinearity $n = 1, 2, 3, 4$.}
\setcounter{equation}{0}
\renewcommand{\theequation}{H\arabic{equation}}

\begin{align}
\mathcal{F} &= 6 \sinh^{2} 2r \\
\mathcal{F} &= 6 \sinh^{2} 2r \left(-20 \cosh 2r+15 \cosh 4r+14\right) \\
\mathcal{F} &=\frac{9}{16} \sinh^{2}r \cosh^{2}r \left(-31320 \cosh 2r +25700 \cosh 4r-17640 \cosh 6r +6895 \cosh 8r +16749\right) \\
\mathcal{F} &=\frac{45}{2} \sinh^{4}r \cosh^{2}r \left(33944 \cosh 2r -29932 \cosh 4r +25921 \cosh 6r-19530 \cosh 8r+11655 \cosh 10r-17258 \right)
\end{align}

\bibliographystyle{apsrev4-1}
\bibliography{Reference}
\end{document}